\documentclass[11pt]{article}
\usepackage{complexity}

\usepackage[english]{babel}
\usepackage[utf8x]{inputenc}
\usepackage[T1]{fontenc}
\usepackage{enumitem}
\usepackage[margin=1.in,marginparwidth=1.75cm]{geometry}

\usepackage[algo2e, ruled, vlined]{algorithm2e}
\usepackage{algorithm}
\usepackage{amsmath}
\usepackage{amsfonts}
\usepackage{graphicx}
\usepackage{amsthm}
\usepackage{amssymb}
\usepackage{dsfont}
\usepackage[noend]{algorithmic} 
\usepackage{bbm}

\usepackage[T1]{fontenc}
\usepackage{microtype}          %
\usepackage[full]{textcomp}

\usepackage{lipsum}
\usepackage{xspace}
\usepackage{relsize}
\usepackage{framed}
\usepackage{xcolor}
\usepackage{graphicx}
\usepackage{multirow}
\usepackage{subcaption}
\usepackage{float}

\usepackage[colorinlistoftodos]{todonotes}
\usepackage[colorlinks=true, allcolors=blue]{hyperref}
\usepackage{cleveref}

\newtheorem{theorem}{Theorem}

\newtheorem{lemma}{Lemma}

\theoremstyle{remark}

\theoremstyle{definition}
\newtheorem{definition}{Definition}

\def\final{0}  
\def\iflong{\iffalse}
\ifnum\final=0  
\newcommand{\knote}[1]{{\color{blue}[{Karthik: \bf #1}]\marginpar{\color{blue}*}}}
\newcommand{\znote}[1]{{\color{purple}[{Weihao: \bf #1}]\marginpar{\color{purple}*}}}
\newcommand{\cnote}[1]{{\color{magenta}[{Chandra: \bf #1}]\marginpar{\color{magenta}*}}}
\else 
\newcommand{\knote}[1]{}
\newcommand{\znote}[1]{}
\newcommand{\cnote}[1]{}
\fi  

\usepackage{float}
\usepackage{booktabs}
\usepackage{multirow}
\usepackage{thmtools} 
\usepackage{thm-restate}
\usepackage{lipsum}

\newcommand{\Z}{\mathbb{Z}}
\renewcommand{\R}{\mathbb{R}}

\newif\ifdraft
\draftfalse

\newcommand{\Ext}{\mathsf{Ext}}

\title{Almost Tight Hardness of Approximation for $0$-Extension}

\begin{document}
\author{Qingyun Chen\thanks{University of California, Santa Cruz. Email: {\tt qingyun.chen152@gmail.com}. Supported in part by NSF grants CCF-2423106, CCF-2537126, and CCF-2535599.}
\and Weihao Zhu\thanks{University of Illinois, Urbana-Champaign. Email: {\tt weihaoz3@illinois.edu}. Supported in part by NSF grant CCF-2402667.}
}
\date{}
\maketitle

\pagenumbering{gobble}
\begin{abstract}
    In the $0$-Extension problem, we are given an undirected graph $G=(V,E)$ with non-negative edge weights $w: E\rightarrow \R_{\geq 0}$, a collection $T \subseteq V$ of $k$ terminals, and a metric $D$ over terminals $T$. The goal is to find a retraction $f:V\rightarrow T$ that maps every terminal to itself, while minimizing the sum over all edges of the product of the edge's weight and the distance in $D$ between the terminals to which its endpoints are assigned, i.e.,
    \[
        \sum_{\{u,v\}\in E} w_{\{u,v\}} \cdot D(f(u), f(v)).
    \]
    Fakcharoenphol, Harrelson, Rao, and Talwar~\cite{fakcharoenphol2003improved} achieve the best known approximation ratio of $O(\log k / \log \log k)$, while the known approximation hardness is $\Omega(\sqrt{\log k})$ assuming the unique games conjecture~\cite{karloff2006earthmover,manokaran2008sdp}.

    In this work, we establish an $\Omega(\log^{1-\varepsilon}k)$ hardness of approximation for every constant $\varepsilon\in (0,1)$ and arbitrarily large fixed $k$, almost matching the upper bound in the exponent of $\log k$. The main technical contribution is an $\Omega(\log^{1-\varepsilon}k)$ integrality gap for the earthmover relaxation. Our construction is inspired by randomized extensions of expanders introduced by Schwartz and Tur~\cite{SchwartzT21} for the metric relaxation. The result then follows from an existing theorem by Manokaran, Naor, Raghavendra, and Schwartz~\cite{manokaran2008sdp} that translates the integrality gap for the earthmover relaxation to the approximation hardness for every fixed $k$.

\end{abstract}

\newpage
\tableofcontents

\newpage
\pagenumbering{arabic}
\section{Introduction}

The classical Multiway Cut problem, introduced by Dahlhaus et al.~\cite{dahlhaus1994complexity}, is defined on an undirected graph with nonnegative edge weights and $k$ designated vertices called \emph{terminals}. The goal is to partition the vertices into $k$ parts, each containing exactly one terminal, while minimizing the total weight of edges between parts. The problem has been studied extensively, with a substantial body of work on approximation algorithms and integrality gaps~\cite{freund2000lower,cualinescu1998improved,brakensiek2026improved,sharma2014multiway,buchbinder2019simple,buchbinder2017simplex,buchbinder2013simplex,berczi2020improving,angelidakis2017improved}. A more general formulation is the Metric Labeling problem, introduced by Kleinberg and Tardos~\cite{kleinberg2002approximation} and motivated by classification problems in machine learning. In this problem, each vertex receives a label from a set equipped with a metric. Assigning a label to a vertex incurs a prescribed nonnegative cost, and each edge incurs its weight times the distance between the labels of its endpoints. The objective is to minimize the sum of these assignment and edge costs. Subsequent work has investigated approximation algorithms and hardness for this more general problem~\cite{chekuri2004linear,chuzhoy2007hardness,karloff2006earthmover,manokaran2008sdp}.

The $0$-Extension problem, introduced by Karzanov~\cite{karzanov1998minimum}, is a fundamental graph partitioning problem that combines the terminal constraints of Multiway Cut with the metric edge costs of Metric Labeling. An instance $\mathcal{I}=(G,w,T,D)$ consists of an undirected graph $G=(V,E)$ with nonnegative edge weights $w:E\to\mathbb{R}_{\geq 0}$, a set $T\subseteq V$ of $k$ terminals, and a metric $D$ on $T$. The goal is to find a retraction $f:V\to T$, where $f(t_i) =t_i$ for every $i \in [k]$, that minimizes
\[
    \sum_{\{u,v\}\in E} w_{\{u,v\}}\cdot D(f(u),f(v)).
\]
We denote the minimum cost by $\operatorname{OPT}(\mathcal{I})$. For the uniform terminal metric, the objective is precisely the weight of edges crossing the partition of terminals, recovering the multiway cut problem. More generally, the $0$-Extension problem is also a special case of the metric labeling problem in which terminals must receive their own labels and non-terminals incur no assignment costs. Along with the development of algorithmic techniques and lower bounds, $0$-Extension presents close connections to other topics, such as Lipschitz extension in Banach spaces \cite{lee2005extending,makarychev2016metric}, metric embeddings \cite{krauthgamer2004measured,fakcharoenphol2003tight}, and vertex sparsifiers \cite{makarychev2016metric,chen2026lower}.

Two linear programming relaxations have been studied for the $0$-Extension problem. The first is the \emph{metric relaxation} introduced by Karzanov~\cite{karzanov1998minimum}. Every feasible assignment $f$ induces distances $d_f(u,v):=D(f(u),f(v))$ on $V$ that agree with $D$ on the terminals. The metric relaxation retains the metric constraints and the prescribed terminal distances, but drops the requirement that the distances arise from an assignment, which can be formalized as follows:
\begin{equation}
\begin{aligned}
    \operatorname{MET}(\mathcal{I}):={}&
    \min_d \sum_{\{u,v\}\in E} w_{\{u,v\}}\cdot d(u,v)\\
    \text{subject to}\quad
    &d(u,v)\leq d(u,z)+d(z,v)
        &&\forall u,v,z\in V,\\
    &d(u,v)=d(v,u)\geq 0
        &&\forall u,v\in V,\\
    &d(v,v)=0
        &&\forall v\in V,\\
    &d(a,b)=D(a,b)
        &&\forall a,b\in T. \notag
\end{aligned}
\end{equation}
C\u{a}linescu, Karloff, and Rabani~\cite{calinescu2005approximation} gave an $O(\log k)$-approximation algorithm by rounding the metric relaxation. Fakcharoenphol, Harrelson, Rao, and Talwar~\cite{fakcharoenphol2003improved} improved the approximation ratio to $O(\log k/\log\log k)$, which remains the best known. For the metric relaxation itself, C\u{a}linescu, Karloff, and Rabani~\cite{calinescu2005approximation} established an $\Omega(\sqrt{\log k})$ integrality gap. Schwartz and Tur~\cite{SchwartzT21} improved this bound to $\Omega(\log^{2/3}k)$ through a novel randomized extensions of expanders.

The second relaxation is the \emph{earthmover relaxation}, introduced by Chekuri, Khanna, Naor, and Zosin~\cite{chekuri2004linear} for metric labeling. It is defined using the \emph{earthmover distance} between probability distributions on the terminal set. Let $\Delta(T)$ denote the set of probability distributions on $T$. For $\mu,\nu\in\Delta(T)$, their earthmover distance $W_D(\mu,\nu)$ is the minimum cost of transporting mass from $\mu$ to $\nu$, where transporting one unit of mass from terminal $a$ to terminal $b$ costs $D(a,b)$. Formally,
\begin{equation}\label{eq:intro-earthmover-distance}
\begin{aligned}
    W_D(\mu,\nu):={}&
    \min_{\pi}\sum_{a,b\in T}\pi(a,b)\cdot D(a,b)\\
    \text{subject to}\quad
    &\sum_{b\in T}\pi(a,b)=\mu(a)
        &&\forall a\in T,\\
    &\sum_{a\in T}\pi(a,b)=\nu(b)
        &&\forall b\in T,\\
    &\pi(a,b)\geq 0
        &&\forall a,b\in T. \notag
\end{aligned}
\end{equation}
Here, $\pi(a,b)$ specifies the mass transported from $a$ to $b$, and the first two constraints ensure that the transported mass has the prescribed source and destination distributions. The earthmover relaxation assigns a probability distribution $\mu_v$ on $T$ to every vertex $v$, while keeping each terminal assigned to itself. It minimizes the weighted sum of earthmover distances between the distributions assigned to adjacent vertices:
\begin{equation}\label{eq:intro-earthmover-lp}
\begin{aligned}
    \operatorname{EM}(\mathcal{I}):={}&
    \min_{\mu}
    \sum_{\{u,v\}\in E}w_{\{u,v\}}\cdot W_D(\mu_u,\mu_v)\\
    \text{subject to}\quad
    &\mu_v\in\Delta(T)
        &&\forall v\in V,\\
    &\mu_a(a)=1
        &&\forall a\in T. \notag
\end{aligned}
\end{equation}
Expanding each earthmover distance using its transportation formulation gives a linear program.

Every integral assignment gives a feasible earthmover solution by concentrating $\mu_v$ at $f(v)$. Moreover, the distances $d(u,v):=W_D(\mu_u,\mu_v)$ induced by any feasible family of distributions satisfy the constraints of the metric relaxation. Consequently,
\[
    \operatorname{MET}(\mathcal{I})
    \leq \operatorname{EM}(\mathcal{I})
    \leq \operatorname{OPT}(\mathcal{I}).
\]
The earthmover relaxation is therefore at least as strong as the metric relaxation, so an integrality gap for the metric relaxation alone does not establish the same gap for the earthmover relaxation.

Karloff, Khot, Mehta, and Rabani~\cite{karloff2006earthmover} proved an $\Omega(\sqrt{\log k})$ integrality gap for the earthmover relaxation. This relaxation is particularly important for approximation hardness. Manokaran, Naor, Raghavendra, and Schwartz~\cite{manokaran2008sdp} showed that, for every fixed $k\geq 1$, a fixed instance with $k$ terminals and earthmover integrality gap $\alpha>1$ yields NP-hardness of approximation within every fixed factor $r$ satisfying $1\leq r<\alpha$, assuming the Unique Games Conjecture. Their reduction preserves the terminal metric and hence the number of terminals. Combined with the gap of Karloff et al.~\cite{karloff2006earthmover}, this gave the previously best known $\Omega(\sqrt{\log k})$ hardness of approximation under the Unique Games Conjecture, leaving a substantial gap from the $O(\log k/\log\log k)$ upper bound.

\subsection{Our Results}

Our main technical result is an almost logarithmic integrality gap for the earthmover relaxation.

\begin{theorem}\label{thm:earthmover-gap}
    For every constant $\varepsilon\in(0,1)$, there exists a constant $c_\varepsilon>0$ such that, for arbitrarily large $k$, there exists a $0$-Extension instance with $k$ terminals whose integrality gap for the earthmover relaxation is at least $c_\varepsilon(\log^{1-\varepsilon}k)$.
\end{theorem}

The reduction of Manokaran, Naor, Raghavendra, and Schwartz~\cite{manokaran2008sdp} converts integrality gaps for the earthmover relaxation into approximation hardness for every fixed $k$ under the Unique Games Conjecture, for which a proof has been claimed very recently~\cite{OAI:The-Unique-Games-Theorem-September-23-2026}. Thus, we obtain the following hardness result, which almost matches the best known $O(\log k/\log\log k)$ approximation ratio in the exponent of $\log k$.

\begin{theorem}\label{thm:nearlog-hardness}
    For every constant $\varepsilon\in(0,1)$, there exists a constant $c_{\varepsilon}>0$ such that, for arbitrarily large fixed integer $k\geq 1$, it is NP-hard to approximate $0$-extension on instances with exactly $k$ terminals within a factor of $c_\varepsilon(\log^{1-\varepsilon}k)$.
\end{theorem}

\subsection{Overview of Our Approach}

\paragraph{Expander towers and the fractional solution.}
Our construction builds on the randomized extension of Schwartz and Tur~\cite{SchwartzT21}. We start with an $n$-vertex constant-degree expander $H$ whose girth is at least $g=\Theta(\log n)$. An extension replaces each vertex by a copy of $H$, called a \emph{core}, and each edge by a uniformly random perfect matching between the corresponding cores. We introduce a pruning step that deletes every matching edge lying on a cycle with fewer than $g$ edges, and show that every matching retains most of its edges. For a fixed integer $t\geq 2$, performing $t-1$ successive pruned extensions gives an expander tower $X$ with $N=n^t$ vertices. Its smallest nonsingleton clouds, called $1$-clouds, are copies of $H$; each larger cloud contains $n$ children connected according to $H$. Building on the local sparsity bounds of Charikar, Makarychev, and Makarychev~\cite{charikar2010local}, we also ensure that sets containing at most $n^{0.9}$ vertices in each core induce graphs of fractional arboricity $1+O(1/\log n)$ at every stage. To construct the instance, we assign edge lengths that increase with the level. For every vertex $x\in V(X)$, we add a distinct terminal $x_T$ and an edge $\{x,x_T\}$, called a \emph{terminal edge}. We inherit the terminal metric from $X$ by setting $D(x_T,y_T)=d_X(x,y)$. Each terminal edge therefore charges for the distance between a vertex's assigned terminal and its own terminal. The fractional solution assigns each nonterminal vertex the uniform distribution on the terminals attached to its $1$-cloud. It pays nothing inside a $1$-cloud, while the surviving matchings transport almost all the mass between neighboring $1$-clouds at small cost. With carefully designed edge weights, the total fractional cost is $O(N)$. We refer readers to \Cref{section:expander-tower,sec:gap-instance} for the construction and fractional analysis. It remains to show that every integral assignment has much larger cost.

\paragraph{Random partitions from local sparsity.}
For the integral analysis, we need to group nearby assigned labels into parts of small diameter. The labels may depend on the entire realized tower, so we require a partition guarantee for every set of at most $n$ vertices after the tower has been fixed. The fractional-arboricity bound provides the starting point: it gives a distribution over forests in which every edge is omitted with probability only $O(1/\log n)$. Randomly cutting these forests produces bounded-diameter parts while rarely separating nearby vertices. An arbitrary set, however, may contain too many vertices in some cores for the sparsity bound to apply directly. We therefore consider short metric neighborhoods of the set and handle cores containing many neighborhood vertices separately. We partition these cores by contracting them to vertices in the preceding stage of the tower, where local sparsity can again be used. The remaining edges admit forest sampling. We then extend the partition of the selected cores through the sampled forest, obtaining the required partition for the original set. We refer readers to \Cref{sec:random-partitions} for further details.

\paragraph{Representatives and the integral lower bound.}
Suppose an integral assignment has small cost. We construct a representative for every cloud and track the vertices whose assigned labels are far from their cloud's representative. In each $1$-cloud, we choose the most frequent assigned label. Expansion and the additional weight on edges inside these clouds bound the total number of exceptions. We then proceed inductively. For a parent cloud, we apply the random partition to the representatives of its $n$ children. When two children have few exceptions, a small cost on the matching between them forces their representatives to be close, so the partition is unlikely to separate them. Thus, matching costs and the existing exceptions bound the expected number of edges of $H$ crossing the induced grouping of children. Expansion bounds the number of children outside the largest group. We choose the parent representative from this group, and the diameter bound ensures that vertices already close to their child representative remain close to the parent representative. Averaging the exceptions over all clouds at each level eventually gives one representative close to the labels of most vertices of $X$. In contrast, projection onto the bounded-degree top core shows that most original vertices are far from this representative. A constant fraction of vertices must therefore be assigned to distant terminals, forcing their terminal edges to have large total cost. Our choice of length scales and weights yields the lower bound $\Omega(N(\log n)^{(t-1)/(t+1)})$, and choosing $t$ sufficiently large in terms of $\varepsilon$ gives the claimed integrality gap. We refer readers to \Cref{sec:integral-lower-bound} for the full argument.

\subsection{Other Related Work}
\paragraph{$0$-Extension.} Better approximation guarantees are available for special cases. For planar graphs, C\u{a}linescu, Karloff, and Rabani \cite{calinescu2005approximation} gave $O(1)$-approximation. $O(\sqrt{\text{diam}(D)})$-approximation exists where $\text{diam}(D)$ is the ratio between the largest and smallest distance in $D$ and $O(1)$-approximation if $D$ is the shortest path metric of a high girth expander \cite{karloff2006earthmover}. Archer et al. \cite{archer2004approximate} gave an $O(\alpha)$-approximation when it is $\alpha$-decomposable, and hence $O(1)$-approximation for planar terminal metrics. In terms of $n$, a hardness of $\Omega(\log^{1/4-\varepsilon} n)$-approximation, for any $\varepsilon > 0$, was given by \cite{karloff2006earthmover}, assuming $\mathsf{NP} \not\subseteq \mathsf{DTIME}(n^{\text{polylog}n})$.

\paragraph{Multiway Cut.} The problem was first introduced by \cite{dahlhaus1994complexity}, together with a simple $2(1-1/k)$-approximation greedy algorithm. It was improved to a $3/2-1/k$-approximation using a geometric relaxation by \cite{cualinescu1998improved}. Later, a sequence of work \cite{buchbinder2017simplex,karger1999rounding,sharma2014multiway,brakensiek2026improved} based on the geometric relaxation culminated in an approximation of $1.2787$. The work of \cite{manokaran2008sdp} also established a translation from integrality gap to approximation hardness for the multiway cut problem assuming the unique games conjecture. \cite{freund2000lower} gave an $8/(7+1/(k-1))$ integrality gap, which was subsequently improved to $1.2$ \cite{angelidakis2017improved} and $1.20016$ \cite{berczi2020improving}.

\paragraph{Metric Labeling.} Kleinberg and Tardos \cite{kleinberg2002approximation} first introduced the problem, motivated by applications in machine learning, and gave an $O(\log k\log \log k)$-approximation algorithm and a $2$-approximation for uniform metric. Later, it was improved to an $O(\log k)$-approximation \cite{fakcharoenphol2003tight}. Chekuri, Khanna, Naor, and Zosin \cite{chekuri2004linear} introduced the earthmover relaxation and gave an $O(\log k)$-approximation rounding algorithm. 
Archer et al. \cite{archer2004approximate} presented an $O(\log n)$-approximation algorithm using tree embeddings, where $n$ is the number of vertices.
\cite{karloff2006earthmover} showed an integrality gap of $\Omega(\log k)$. This result can be converted into a hardness of $\Omega(\log k)$-approximation assuming the unique games conjecture by $\cite{manokaran2008sdp}$. If $D$ is a truncated linear metric, it is a $4$-approximation \cite{gupta2000constant} and improved to $2+\sqrt{2}$ \cite{chekuri2004linear}. Further, if $D$ is a linear metric, it has an exact solution \cite{chekuri2004linear}. 

\paragraph{Vertex Sparsifiers. } Vertex sparsifiers compress a graph while approximately preserving cuts or multicommodity flows among a given set of $k$ terminals. Their connection to $0$-Extension arises from viewing a terminal assignment as a contraction: vertices assigned to the same terminal are identified, and distributions over such contractions provide a natural approach to sparsification. Makarychev and Makarychev \cite{makarychev2016metric} formalized connections among vertex sparsifiers, metric-extension operators, and Lipschitz extendability, and gave polynomial-time constructions of terminal-only cut and flow sparsifiers with quality $O(\log k/\log\log k)$. They also established an $\Omega(\sqrt{\log k/\log\log k})$ lower bound for terminal-only flow sparsifiers. Allowing additional Steiner vertices introduces a tradeoff between sparsifier size and quality. In this direction, Chen and Tan \cite{chen_et_al:LIPIcs.ICALP.2024.47} showed that a Steiner-node variant of $0$-Extension admits a $\Omega(\log\log k)$ integrality gap for the metric relaxation even when a superlinear $O(k \log^{1-\varepsilon} k)$ number of Steiner vertices is allowed. More recently, Chen, Tan, and Yang \cite{chen2026lower} proved polylogarithmic $\Omega((\log k)^{0.3})$ quality lower bounds for contraction-based flow sparsifiers, even when the number of permitted Steiner vertices exceeds $k2^{(\log k)^{\Omega(1)}}$.

\section{Preliminaries}\label{section:prelim}

Let $G=(V,E)$ be an undirected graph with positive edge lengths $\ell:E\to \R_+$. For every pair of vertices $x,y\in V$, let $d_G(x,y)$ be the length of a shortest path connecting $x$ and $y$ in graph $G$, with $d_G(x,y)=\infty$ if no such path exists. For every subset $S\subseteq V$ of vertices, let $G[S]$ be the induced subgraph on $S$, and $\delta_G(S)$ be the collection of edges with exactly one endpoint in $S$. We call graph $G$ \emph{simple} if it has neither self-loops nor parallel edges. For every integer $r\geq 1$, we call graph $G$ \emph{$r$-regular} if every vertex has degree exactly $r$.

The \emph{diameter} of $G$ is $\max_{x,y\in V}d_G(x,y)$, where distances are measured using the specified edge lengths. The \emph{unweighted diameter} of $G$ is the maximum, over all pairs of vertices $x,y \in V$, of the minimum number of edges in a path connecting $x$ and $y$. The \emph{girth} of $G$ is the minimum number of edges in a cycle, with girth $\infty$ if the graph $G$ is acyclic. We emphasize that both unweighted diameter and girth are unrelated to the edge weights. We define the \emph{edge expansion} of $G$ as
\[
    h(G):=\min\left\{\frac{|\delta_G(S)|}{|S|}:S\subseteq V, 0<|S|\leq |V|/2\right\}
\]
and the \emph{fractional arboricity} of $G$ as
\[
    \Gamma(G):=\max\left\{1,\max\left\{\frac{|E(G[S])|}{|S|-1}:S\subseteq V, |S|\geq 2\right\}\right\},
\]
where $|E(G[S])|$ is the number of edges in the subgraph induced by $S$.

For every integer $n\in \Z_+$, we let $[n]$ denote the set $\{1,2,\ldots, n\}$. All logarithms are by default natural throughout this work.

\section{Expander Towers}\label{section:expander-tower}

In this section, we construct the expander towers used in our integrality gap instances. Throughout this section, we fix an integer $t\geq 1$, which specifies the number of levels in the expander tower. \Cref{subsec:base-expander} provides the base expanders, whose small induced subgraphs have fractional arboricity close to one. In \Cref{subsec:expander-extension}, we introduce the randomized extension of Schwartz and Tur~\cite{SchwartzT21}. We then remove short cycles by a pruning step and establish the properties of the pruned extension in \Cref{subsec:expander-pruning}. We iterate this construction to obtain expander towers in \Cref{subsec:expander-tower} and establish the distance and matching properties needed for the integrality gap analysis in \Cref{subsec:tower-property}.

\subsection{Expanders with Local Small Fractional Arboricity}\label{subsec:base-expander}

In this section, we construct the base expander used at every level of the tower. We require its small induced subgraphs to have fractional arboricity close to one. Our starting point is the following expander result.

\begin{lemma}[\cite{arora2002proving,arora2006local,charikar2010local}]\label{lemma:previous-expander-result}
    There are constants $0<h_0\leq 1$, $\gamma^*>0$, and $a>0$ such that, for every sufficiently large even integer $n$ and every integer $k\in [n-1]$, there exists a simple $3$-regular graph $H$ on $n$ vertices with edge expansion at least $h_0$ and girth at least $\gamma^*\log n$, satisfying
    \[
        |E(H[S])|\leq \left(1+\frac{a}{\log(n/k)}\right)|S|
    \]
    for every $S\subseteq V(H)$ with $|S|\leq k$.
\end{lemma}

The following lemma collects the properties of the base expander needed in later sections.

\begin{lemma}\label{lemma:fa-expander}
    There are constants $0<h_0\leq 1$, $\kappa>0$, and
    \[
        0<\gamma \leq \frac{1}{40000t\log (40t)}
    \]
    such that, for every sufficiently large even integer $n$, there exists a simple graph $H$ on $n$ vertices satisfying the following properties:
    \begin{enumerate}
        \item the maximum degree is at most $d=3$;
        \item the edge expansion is at least $h_0$;
        \item the girth is at least $\lfloor \gamma \log n \rfloor$;
        \item the unweighted diameter is at most $\eta\log n$, where $\eta:=50/h_0$;
        \item for every $S\subseteq V(H)$ with $|S|\leq n^{0.9}$,
        \[
            \Gamma(H[S])\leq 1+\frac{\kappa}{\log n}.
        \]
    \end{enumerate}
\end{lemma}
\begin{proof}
    We let $h_0$, $\gamma^*$, and $a$ be the absolute constants in \Cref{lemma:previous-expander-result}, and set
    \[
        \gamma:=\min\left\{\gamma^*, \frac{1}{40000t\log (40t)}\right\} \qquad \text{and} \qquad \kappa:=10a+\frac{4}{\gamma}.
    \]
    For a sufficiently large even integer $n$, we apply \Cref{lemma:previous-expander-result} with $k:=\lfloor n^{0.9}\rfloor$ and let $H$ be the resulting graph. The first three properties follow immediately. We next verify the unweighted diameter and fractional arboricity bounds.

    For every vertex $u\in V(H)$ and integer $r\geq 0$, let $B_r(u)$ denote the set of vertices reachable from $u$ by a path with at most $r$ edges. Whenever $|B_r(u)|\leq n/2$, the edge expansion and degree bounds imply that
    \[
        |B_{r+1}(u)| \geq |B_r(u)|+\frac{|\delta_H(B_r(u))|}{d}  \geq \left(1+\frac{h_0}{d}\right) |B_r(u)|.
    \]
    We set $R:=\left\lceil \frac{20}{h_0}\log n\right\rceil$ and show that $|B_R(u)|>n/2$. Otherwise, for every $0\leq r < R$, we have $|B_r(u)|\leq n/2$. Iterating the preceding inequality gives
    \[
        |B_R(u)|\geq \left(1+\frac{h_0}{d}\right)^R\geq n>n/2,
    \]
    which leads to a contradiction. Hence, $|B_R(u)|>n/2$. This implies that for arbitrary two vertices $u,v \in V(H)$, the sets $B_R(u)$ and $B_R(v)$ intersect. Thus, the unweighted diameter of $H$ is at most
    \[
        2R\leq \frac{50}{h_0}\log n =\eta \log n.
    \]

    It remains to bound the fractional arboricity of small induced subgraphs. We fix a set $S\subseteq V(H)$ of vertices with $|S|\leq n^{0.9}$. For every $U\subseteq S$ with $|U|\geq 2$, \Cref{lemma:previous-expander-result} shows that
    \begin{align}
        |E(H[U])|\leq \left(1+\frac{a}{\log (n/k)}\right) |U| \leq \left(1+\frac{10a}{\log n}\right)|U|, \label{inequality:fa-expender-edge-bound}
    \end{align}
    where the last inequality is because $k=\lfloor n^{0.9} \rfloor$. If $|U|<\lfloor \gamma \log n \rfloor$, then $H[U]$ is a forest since the girth of $H$ is at least $\lfloor \gamma \log n\rfloor$ by the third property, which further implies that $|E(H[U])|\leq |U|-1$. Otherwise, for sufficiently large $n$, we have
    $$\begin{aligned}
        \frac{|E(H[U])|}{|U|-1} &\leq \left(1+\frac{10a}{\log n}\right)\left(1+\frac{1}{|U|-1}\right) && \text{(by inequality~(\ref{inequality:fa-expender-edge-bound}))} \\
        &= 1+\frac{10a}{\log n} + \left(1+\frac{10a}{\log n}\right)\frac{1}{|U|-1} \\
        & \leq 1+\frac{10a}{\log n} + \frac{2}{|U|-1} && \text{(since $1+\frac{10a}{\log n}\leq 2$)} \\
        & \leq 1+\frac{10a}{\log n} + \frac{2}{(\gamma/2)\log n} && \text{(since $|U|-1\geq (\gamma/2)\log n$)} \\
        & = 1 + \frac{\kappa}{\log n},
    \end{aligned}$$
    where the last equality is because $\kappa=10a+4/\gamma$. Therefore,
    \[
        \Gamma(H[S])\leq 1+\frac{\kappa}{\log n}.
    \]
\end{proof}

For the remaining sections of \Cref{section:expander-tower}, we fix a sufficiently large even integer $n$.

\subsection{Randomized Extension}\label{subsec:expander-extension}

\begin{figure}[ht]
    \centering
    \includegraphics[width=0.75\linewidth]{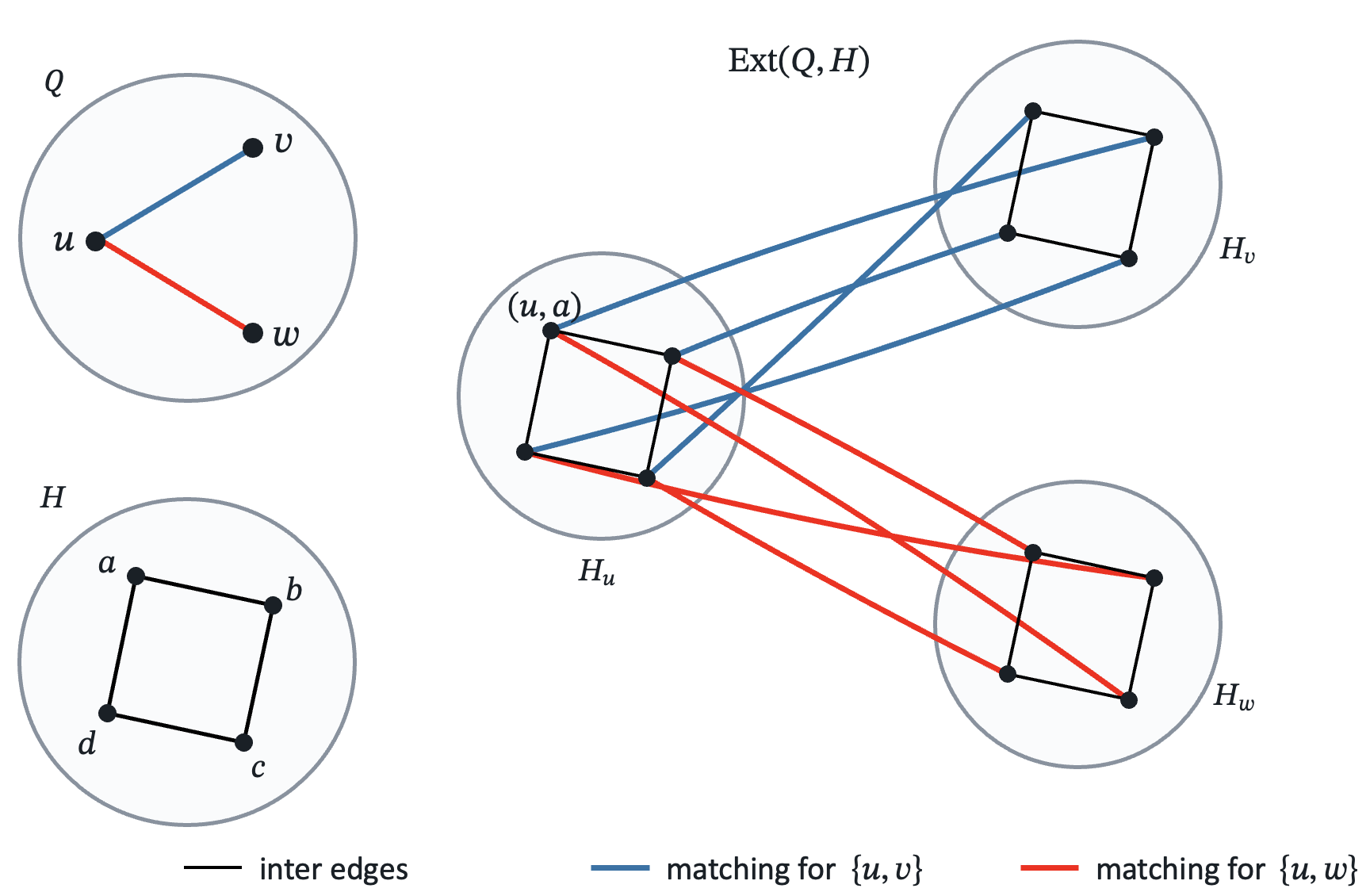}
    \caption{The figure shows two undirected graphs $Q$ and $H$, together with a realization of $\Ext(Q,H)$ on the right side. Every vertex of $Q$ is replaced by a copy of $H$, and each edge is replaced by an independently chosen uniformly random perfect matching between the corresponding copies. Black edges are inter edges, while blue and red edges form the two outer matchings.
    }
    \label{figure:random-extension}
\end{figure}

Our tower construction builds on the randomized extension of Schwartz and Tur~\cite{SchwartzT21}. Informally, for undirected graphs $Q$ and $H$, the randomized extension of $Q$ by $H$ is a random graph obtained by replacing each vertex of $Q$ with a copy of $H$ and each edge of $Q$ with an independently chosen uniformly random perfect matching between the corresponding copies. We formalize it as follows. See an example in \Cref{figure:random-extension}.

\begin{definition}[Randomized Extension]
    Let $Q$ and $H$ be finite undirected graphs, and assume that $Q$ has no self-loops. The \emph{randomized extension} of $Q$ by $H$, denoted by $\Ext(Q,H)$, is a random graph with vertex set $V(Q)\times V(H)$ and edge set $E^*$ defined as follows:
    \begin{enumerate}
        \item for every $u\in V(Q)$ and $\{a,b\}\in E(H)$, we include the edge $\{(u,a),(u,b)\}$ in $E^*$;
        \item for every edge $e=\{u,v\}\in E(Q)$, we fix an ordering of the endpoints $(u,v)$, independently choose a uniformly random bijection $\varphi_e:V(H)\to V(H)$, and include in $E^*$ the matching 
        \[
            \left\{\{(u,a),(v,\varphi_e(a))\}:a\in V(H)\right\}
        \]
    \end{enumerate}
\end{definition}

Let $Y=\Ext(Q,H)$ with $|V(H)|=n$. For every $u\in V(Q)$, let $H_u$ be the induced subgraph of $Y$ on $\{(u,a):a\in V(H)\}$. We call $H_u$ a \emph{core}, which is isomorphic to $H$. An edge $e\in E(Y)$ is called an \emph{inter edge} if it belongs to a core and an \emph{outer edge} otherwise. For every edge $e=\{u,v\}\in E(Q)$, we call the perfect matching between $H_u$ and $H_v$ generated by $e$ an \emph{outer matching}.

We refer to $u$ as the \emph{contracted vertex} corresponding to $H_u$, and to $Q$ as the \emph{contracted graph} of $Y$. For an edge $e=\{u,v\}\in E(Q)$, a \emph{lift} of $e$ is an outer edge in the matching between $H_u$ and $H_v$ generated by $e$. More generally, for a path $P=(u_0, \ldots, u_s)$ in $Q$, a lift of $P$ is a path $(x_0,\ldots, x_s)$ in $Y$ such that $x_i\in V(H_{u_i})$ for every $0\leq i \leq s$ and $\{x_{i-1},x_i\}$ lifts the $i$-th edge of $P$ for every $i\in [s]$. Since each matching is perfect, we emphasize that specifying a starting vertex $x_0\in V(H_{u_0})$ uniquely determines the lift of a given path.

We call a set $S\subseteq V(Y)$ \emph{core-sparse} if for every $u\in V(Q)$,
\[
    |S\cap V(H_u)|\leq n^{0.9}.
\]
We next bound the number of outer edges induced by a core-sparse set $S$. To achieve this, we examine the random matchings by revealing their edges one at a time. A \emph{query} is a pair $(e,x)$, where $e=\{u,v\}\in E(Q)$ has fixed ordered endpoints $(u,v)$ and $x\in V(H_u)\cup V(H_v)$. For the bijection $\varphi_e:V(H)\to V(H)$ used in the construction, a query at $x=(u,a)$ returns $(v,\varphi_e(a))$, whereas a query at $x=(v,b)$ returns $(u,\varphi_e^{-1}(b))$. In either case, the answer reveals the matching edge incident to $x$. The \emph{query history} consists of the preceding queries and their answers. Each subsequent query may be chosen using this history. If the selected matching edge has already been revealed, the query provides no new information.

For a query from a vertex of $S$ to a core $H_v$, the revealed edge belongs to $Y[S]$ precisely when its other endpoint lies in $S\cap V(H_v)$. Core-sparsity bounds the size of this target set by $n^{0.9}$. The following lemma gives a conditional probability bound for a general target set $B\subseteq V(H_v)$ determined by the query history.

\begin{lemma}\label{lemma:query-independent}
    Let $Q$ and $H$ be two undirected graphs with $|V(H)|=n$, and let $Y=\Ext(Q,H)$. We condition on a query history $\mathcal{H}$, and consider a query $(e,x)$ whose matching edge has not yet been revealed. Let $H_v$ be the core containing its answer $y$, and let $r$ be the number of previously revealed edges in the matching associated with $e$. Then, for every set $B\subseteq V(H_v)$,
    \[
        \Pr[y\in B \mid \mathcal{H}] \leq \frac{|B|}{n-r}.
    \]
\end{lemma}
\begin{proof}
    Since each query is chosen using only preceding answers, the history $\mathcal{H}$ imposes no restrictions on the selected matching beyond its $r$ revealed pairs. Let $U\subseteq V(H_v)$ be the vertices not incident to these reveal edges. Hence, $|U|=n-r$.

    Conditioned on $\mathcal{H}$, the remaining $n-r$ vertices in each core are matched uniformly at random. Thus, the newly revealed endpoint $y$ is uniformly on $U$, and
    \[
        \Pr[y\in B\mid \mathcal{H}] = \frac{|B\cap U|}{n-r} \leq \frac{|B|}{n-r}.
    \]
\end{proof}

We now introduce local edge labels to describe explorations of a randomized extension, which will be used in the later proof. We assume that $H$ and $Q$ are two undirected graphs with maximum degrees at most $3$ and $10(t-1)$, respectively. We fix an ordering of the edges in $H$ and an ordering of the edges in $Q$. At every vertex $(u,a)\in V(Y)$, we label its incident inter edges using labels in $[3]$, according to the relative order of the edges incident to $a$ in $H$. Similarly, we label its incident outer edges using labels in $\{4, \ldots, 10t\}$, according to the relative order of the edges incident to $u$ in $Q$. We emphasize that some labels may be unused. We call these labels \emph{ports}. An inter port specifies a neighbor within the core, whereas an outer port specifies a matching edge to query. Thus, all port labels are fixed before the random matchings are sampled.

We next combine \Cref{lemma:query-independent} with the local fractional-arboricity property from \Cref{lemma:fa-expander} to obtain a bound on the number of induced edges that holds simultaneously for all sufficiently large connected core-sparse sets.

\begin{lemma}\label{lemma:et-cs}
    Let $H$, $\gamma$, and $\kappa$ be as in \Cref{lemma:fa-expander}, and let $Q$ be a simple graph with at most $n^{t-1}$ vertices and maximum degree at most $10(t-1)$. Let $Y=\Ext(Q,H)$ be the randomized extension of $Q$ by $H$. Then, there exists a constant $\kappa'>0$ depending only on $t$ such that, with probability at least $1-2n^{-4}$, every core-sparse set $S\subseteq V(Y)$ such that $Y[S]$ is connected and $|S|\geq \lfloor \gamma\log n \rfloor$ satisfies
    \[
        |E(Y[S])|\leq \left(1+\frac{\kappa'}{\log n}\right)|S|.
    \]
\end{lemma}
\begin{proof}
    Let $g:=\lfloor \gamma \log n \rfloor$. According to \Cref{lemma:fa-expander}, the girth of $H$ is at least $g$. We also observe that the graph $Y$ has at most $n^t$ vertices and maximum degree at most $10t$. We select $\kappa'$ satisfying
    \[
        \kappa'>\kappa+400t\log (40t)+\frac{20(t+4)}{\gamma},
    \]
    which is a constant only depending on $t$.

    Suppose that a connected core-sparse set $S$ of size $w\geq g$ violates the claimed bound. Consider the subgraph on $S$ consisting of its inter edges. Let $c$ be its number of connected components and $m$ be the number of edges. We observe that each component lies in a single core and contains at most $n^{0.9}$ vertices. Summing the fractional-arboricity bound from \Cref{lemma:fa-expander} over these components gives
    \[
        m\leq \left(1+\frac{\kappa}{\log n}\right)(w-c).
    \]

    We now select a spanning tree for each component. Since $Y[S]$ is connected, these trees can be joined by $c-1$ outer edges to form a spanning tree $T$ of $Y[S]$. Let $b$ be the number of outer edges of $Y[S]$ outside $T$. Therefore,
    $$\begin{aligned}
        b &= |E(Y[S])| - m - (c-1) \\
        &> \left(1+\frac{\kappa'}{\log n}\right) w - \left(1+\frac{\kappa}{\log n}\right) (w-c)-(c-1) \\
        &>\frac{\kappa'-\kappa}{\log n}w.
    \end{aligned}$$

    We next bound the probability that such a set exists by encoding its spanning tree $T$ and the $b$ outer edges outside $T$. To encode $T$, we record its root in $V(Y)$, its rooted ordered tree shape, and the port at the parent of each tree edge. For a fixed realization of $Y$, this information determines all vertices and edges of $T$. We encode each outer edge of $Y[S]$ outside $T$ by marking its port at exactly one endpoint. Each mark specifies a vertex of the recorded tree and a port label at that vertex. Thus, every set violating the claimed bound admits an encoding with $b>(\kappa'-\kappa)w/\log n$ marked ports.

    We first count the number of possible encodings. There are at most $n^t$ choices for the root and $4^{w-1}$ rooted ordered tree shapes on $w$ vertices. Each tree edge has at most $10t$ possible port labels. Each of the at most $10tw$ possible outer ports is either marked or unmarked, giving at most $2^{10tw}$ choices for the marked set. Hence, the number of encodings is at most
    \[
        n^t4^{w-1}(10t)^{w-1}2^{10tw}=n^t(40t)^{w-1}2^{10tw}.
    \]

    We next bound the probability that a fixed encoding with $b>(\kappa'-\kappa)w/\log n$ marked ports is valid. We first explore its labeled tree in depth-first order, revealing a matching edge whenever an outer port is used. We discard the encoding if a specified port is unused, a vertex is repeated, or the resulting vertex set is not core-sparse. Otherwise, the exploration determines a core-sparse set $S$ of size $w$ and a spanning tree $T$ on $S$. We next query the marked ports in a fixed order. The encoding is valid if every marked query reveals a previously unrevealed matching edge whose other endpoint belongs to $S$. In that case, the marked queries reveal $b$ distinct outer edges of $Y[S]$ outside $T$. Every encoding obtained from a violating set is valid in this sense.

    We condition on an arbitrary query history produced by a tree exploration that has not been discarded, which fixes the core-sparse set $S$ and the spanning tree $T$. Consider a subsequent history $\mathcal H$ in which every marked query made so far has revealed a previously unrevealed outer edge of $Y[S]$ outside $T$. If the next marked port is unused or its matching edge has already been revealed, the encoding is invalid. Otherwise, suppose that this query starts in $H_u$ and reveals an endpoint $y\in V(H_v)$. Let $r$ be the number of previously revealed edges in the selected matching. Every such edge has both endpoints in $S$, and these edges are pairwise vertex-disjoint. Hence,
    \[
        r\leq \min\bigl\{|S\cap V(H_u)|,\ |S\cap V(H_v)|\bigr\} \leq n^{0.9}.
    \]
    Applying \Cref{lemma:query-independent} with $B=S\cap V(H_v)$ gives
    \[
        \Pr[y\in S\cap V(H_v)\mid\mathcal H]
        \leq \frac{|S\cap V(H_v)|}{n-r}
        \leq \frac{n^{0.9}}{n-n^{0.9}}
        \leq 2n^{-0.1}
    \]
    for sufficiently large $n$.

    Thus, conditioned on any history in which all previously processed marked ports satisfy the validity condition, the next marked query satisfies that condition with probability at most $2n^{-0.1}$. Repeated conditioning bounds the probability that all $b$ marked queries satisfy this condition by $(2n^{-0.1})^b$. This bound holds for every tree query history for which the encoding has not been discarded. Since $b\leq 10tw$ and $b>(\kappa'-\kappa)w/\log n$, we obtain
    \[
    \begin{aligned}
        \Pr[\text{the fixed encoding is valid}]
        &\leq (2n^{-0.1})^b\\
        &\leq 2^{10tw} \exp\left(-\frac{\kappa'-\kappa}{10}\,w\right).
    \end{aligned}
    \]
    Applying a union bound over all encodings with more than $(\kappa'-\kappa)w/\log n$ marked ports gives
    \[
    \begin{aligned}
        \Pr[\text{a violating set of size }w\text{ exists}]
        &\leq n^t(40t)^{w-1}2^{20tw}
            \exp\left(-\frac{\kappa'-\kappa}{10}\,w\right)\\
        &\leq n^t
            \exp\left(-\frac{2(t+4)}{\gamma}\,w\right),
    \end{aligned}
    \]
    where the last inequality follows from our choice of $\kappa'$ and
    \[
        \log(40t)+20t\log 2\leq 40t\log(40t).
    \]

    Finally, we recall that $g\geq(\gamma/2)\log n$ for sufficiently large $n$. Summing over all $w\geq g$ gives
    \[
    \begin{aligned}
        \Pr[\text{a violating set of size at least }g\text{ exists}]
        &\leq \sum_{w\geq g} n^t\exp\left(-\frac{2(t+4)}{\gamma}\,w\right)\\
        &\leq 2n^t \exp\left(-\frac{2(t+4)}{\gamma}\,g\right) \\
        &\leq 2n^{-4}.
    \end{aligned}
    \]
\end{proof}

For the remaining sections of \Cref{section:expander-tower}, we fix an $n$-vertex expander $H$ as in \Cref{lemma:fa-expander}, with the constants $h_0$, $\gamma$, and $\kappa$ specified there. We will use $H$ as the core in every step of the expander tower construction.

\subsection{Pruning Short Cycles}\label{subsec:expander-pruning}

Let $g:=\lfloor \gamma \log n \rfloor$. We now introduce a pruning step to remove short cycles containing fewer than $g$ edges in the randomized extension by deleting outer edges.

\begin{definition}[Pruned Extension]
    Let $Q$ be a finite undirected graph without self-loops, and let $Y:=\Ext(Q,H)$ be the extension of $Q$ by $H$. The \emph{pruned extension} of $Q$ by $H$, denoted by $\widehat{\Ext}(Q,H)$, is obtained from $Y$ by deleting every outer edge that lies on a cycle of $Y$ with fewer than $g$ edges. We call the remaining outer edges \emph{surviving outer edges}.
\end{definition}

Let $\widehat{Y}:=\widehat{\Ext}(Q,H)$. According to \Cref{lemma:fa-expander}, the graph $H$ has girth at least $g$. This implies that every cycle of $Y$ with fewer than $g$ edges contains at least an outer edge. Therefore, the pruning step removes all such cycles and $\widehat{Y}$ has girth at least $g$. The following lemma shows that this deletion removes few edges from each outer matching.

\begin{lemma}\label{lemma:pruning-loss}
    Let $Q$ be a simple graph with at most $n^{t-1}$ vertices and maximum degree at most $10(t-1)$. Let $Y:=\Ext(Q,H)$ be the randomized extension of $Q$ by $H$, and let $\widehat Y:=\widehat{\Ext}(Q,H)$ be its pruned extension. Then, with probability at least $1-1/n$, at most $\lceil n^{0.99}\rceil$ edges are deleted from every outer matching of $Y$ simultaneously, i.e.,
    \[
        \Pr\left[|M\setminus E(\widehat{Y})|\leq \lceil {n^{0.99}\rceil} \text{ for every outer matching $M$ of $Y$}\right]\geq 1-\frac{1}{n}.
    \]
\end{lemma}

\begin{proof}
    Let $D:=10t$. We observe that the maximum degree of $Y$ is bounded by $D$. We fix an outer matching $M$ between cores $H_u$ and $H_v$ of $Y$, and let $Z\subseteq M$ be the set of edges deleted by pruning. We first show that
    \[
        \Pr\left[|Z|>n^{0.99}\right]\leq \exp(-n^{0.9}),
    \]
    and then take a union bound over all outer matchings in $Y$.

    Suppose that $|Z|>n^{0.99}$. For each deleted edge $e\in Z$, we choose a cycle $C_e$ containing $e$ with fewer than $g$ edges. We root $C_e$ at the endpoint of $e$ in $H_u$. These roots are distinct since $M$ is a perfect matching. We claim that $\{C_e:e\in Z\}$ contains many vertex-disjoint cycles. We fix one of the chosen cycles $C$. If another chosen cycle intersects $C$, its root can be joined to a vertex of $C$ by a path of fewer than $g$ edges. The degree bound implies that the number of possible roots of such cycles is at most
    \[
        |V(C)|\left(\sum_{i=0}^{g-1}D^i\right) < 2gD^g.
    \]
    More precisely, there are fewer than $2gD^g$ chosen cycles intersecting $C$. We repeatedly select a cycle and discard all chosen cycles that intersect it. This procedure produces at least
    \[
        q:=\left\lfloor\frac{n^{0.99}}{2gD^g}\right\rfloor
    \]
    vertex-disjoint cycles. According to \Cref{lemma:fa-expander}, $\gamma \leq \frac{1}{40000t\log (40t)}$, which implies that
    \[
        D^g\leq \left(10t\right)^{\gamma \log n}\leq n^{1/40000}.
    \]
    Consequently, for sufficiently large $n$, we have $q\geq n^{0.98}$, $qg<n/2$, and $4egD^{2g}\leq n^{0.01}$.

    We next bound the probability that such a collection of $q$ cycles exists. We encode these cycles using the ports defined in \Cref{subsec:expander-extension}. First, we record the $q$ distinct roots in $H_u$. For each cycle, we start at its root and traverse its edge in $M$. We then record its length $\ell<g$ and the ports used for its remaining $\ell-1$ edges. For a fixed realization of the randomized extension $Y$, the first edge of the cycle is determined by the root and the matching $M$, and each recorded port determines the next edge. Thus, a valid encoding uniquely determines the cycle.
    
    We first count the number of possible encodings. There are $\binom{n}{q}$ choices for the roots. For each root, there are at most $g$ choices for the cycle length and at most $D$ choices for each port, giving at most $gD^g$ possible encodings. Thus, the number of encodings of the entire collection is at most
    \[
        \binom{n}{q}\cdot (gD^g)^q.
    \]
    We call an encoding \emph{valid} if its prescribed traversals form $q$ vertex-disjoint simple cycles of length less than $g$. The preceding argument shows that a valid encoding exists whenever $|Z|>n^{0.99}$.

    We next bound the probability that a fixed encoding is valid. For a given encoding, we explore its traversals one at a time, processing the roots according to a fixed ordering of $V(H_u)$. We reveal a matching edge whenever an outer port is used. We discard the encoding if a port is unused, a traversal visits a vertex of an earlier cycle, or a traversal repeats a vertex except for its final return to its own root. We also discard the encoding if a traversal does not return to its root after the prescribed number of steps. We bound the probability that the current traversal forms a valid cycle, conditioned on the queries used to reveal the preceding cycles. The inter and outer port labels determine the last outer step of the traversal. For the traversal to be valid, the matching edge used at this step must be previously unrevealed. Otherwise, this edge either belongs to an earlier cycle or has already been traversed in the current traversal. The former case violates vertex-disjointness, and the latter repeats an edge, so neither can occur in a valid encoding.

    Let $x\in V(H_u)$ be the root of the current traversal. After the last outer-edge query, all remaining steps use inter edges. To close the cycle, this query must therefore return to $H_u$ at a vertex that can be joined to $x$ by at most $g$ inter edges. Let $B\subseteq V(H_u)$ be the set of such vertices. This set is fixed once $x$ is specified, and the degree bound gives
    \[
        |B|\leq \sum_{i=0}^{g}D^i\leq 2D^g.
    \]

    If the query has a destination core other than $H_u$, the encoding is invalid. Otherwise, let $z\in V(H_u)$ be its answer, and condition on any preceding query history $\mathcal H$ for which the encoding has not been discarded. Let $r$ be the number of previously revealed edges in the selected matching. Since we explore at most $q$ traversals, each with fewer than $g$ steps, fewer than $qg$ matching edges have been revealed in total. In particular, $r\leq qg$. Applying \Cref{lemma:query-independent} gives
    \[
        \Pr[z\in B\mid\mathcal H]\leq \frac{|B|}{n-r}\leq \frac{2D^g}{n-qg}\leq \frac{4D^g}{n}.
    \]

    Returning an endpoint in $B$ is necessary for the current traversal to form a valid cycle. This probability bound holds for every possible history before its last outer edge query. Hence, conditioning on arbitrary history of the preceding valid traversals, the current traversal is valid with probability at most $4D^g/n$. Applying this bound to the $q$ traversals in order gives
    \[
        \Pr[\text{the fixed encoding is valid}] \leq \left(\frac{4D^g}{n}\right)^q.
    \]
    We recall that a valid encoding exists when $|Z|>n^{0.99}$. Thus, applying a union bound yields
    \[
    \begin{aligned}
        \Pr\left[|Z|>n^{0.99}\right]
        &\leq \binom{n}{q}(gD^g)^q \left(\frac{4D^g}{n}\right)^q\leq \left(\frac{4\mathrm e\,gD^{2g}}{q}\right)^q \leq \exp(-n^{0.9}),
    \end{aligned}
    \]
    where the last inequality is because $4\mathrm e\,gD^{2g}\leq n^{0.01}$ and $q\geq n^{0.98}$.

    Finally, there is one outer matching for each edge of $Q$, and
    \[
        |E(Q)|\leq 5(t-1)n^{t-1}\leq 5tn^{t-1}.
    \]
    Therefore, by a union bound, the probability that there exists an outer matching losing more than $\lceil n^{0.99}\rceil$ edges is at most
    \[
        5tn^{t-1}\exp(-n^{0.9})\leq \frac1n
    \]
    for sufficiently large $n$.
\end{proof}

We now combine \Cref{lemma:et-cs,lemma:pruning-loss} to obtain the properties needed for one step of the tower construction. After pruning, connected induced subgraphs with fewer than $g$ vertices are trees, while \Cref{lemma:et-cs} bounds the number of edges induced by larger connected core-sparse sets. Together, these guarantees yield a fractional-arboricity bound for every core-sparse induced subgraph.

\begin{lemma}\label{lemma:pruned-extension}
    Let $Q$ be a simple graph with at most $n^{t-1}$ vertices and maximum degree at most $10(t-1)$. There exists a constant $\kappa_{\mathrm{ext}}>0$ depending only on $t$ and a realization $Y$ of $\Ext(Q,H)$ such that its pruned extension $\widehat Y$ satisfies the following properties:
    \begin{enumerate}
        \item every outer matching of $Y$ retains at least $n-\lceil n^{0.99}\rceil$ edges in $\widehat Y$;
        \item the girth of $\widehat Y$ is at least $g$;
        \item for every core-sparse set $S\subseteq V(\widehat Y)$,
        \[
            \Gamma(\widehat Y[S])
            \leq 1+\frac{\kappa_{\mathrm{ext}}}{\log n}.
        \]
    \end{enumerate}
\end{lemma}

\begin{proof}
    Let $\kappa'$ be the constant in \Cref{lemma:et-cs}, and set
    \[
        \kappa_{\mathrm{ext}}:=\kappa'+\frac{8}{\gamma}.
    \]
    For sufficiently large $n$, the conclusions of \Cref{lemma:et-cs,lemma:pruning-loss} hold simultaneously with probability at least
    \[
        1-2n^{-4}-\frac1n>0.
    \]
    We fix such a realization of $Y$ and prune it to obtain $\widehat Y$. The first property follows from \Cref{lemma:pruning-loss}. Since every core has girth at least $g$, every shorter cycle in $Y$ contains an outer edge and is removed by pruning. Thus, the second property also holds.

    It remains to prove the fractional-arboricity bound. We first consider a core-sparse set $U$ such that $\widehat Y[U]$ is connected and $|U|\geq 2$. If $|U|<g$, then $\widehat Y[U]$ is a tree, so $|E(\widehat Y[U])|=|U|-1$. Otherwise, \Cref{lemma:et-cs} gives
    \[
        |E(\widehat Y[U])| \leq |E(Y[U])| \leq \left(1+\frac{\kappa'}{\log n}\right)|U|.
    \]
    Consequently,
    \[
    \begin{aligned}
        \frac{|E(\widehat Y[U])|}{|U|-1} &\leq \left(1+\frac{\kappa'}{\log n}\right)\left(1+\frac{1}{|U|-1}\right)\\
        &=1+\frac{\kappa'}{\log n}+\left(1+\frac{\kappa'}{\log n}\right)\frac{1}{|U|-1}\\
        &\leq1+\frac{\kappa'}{\log n}+\frac{2}{g-1} \ \ \ \ \ \text{(since $1+\frac{\kappa'}{\log n}\leq 2$ and $|U|\geq g$)}\\
        &\leq1+\frac{\kappa_{\mathrm{ext}}}{\log n}. \ \ \ \ \ \text{(since $g-1\geq (\gamma/2)\log n$)}
    \end{aligned}
    \]

    Now we fix an arbitrary core-sparse set $S$. To bound $\Gamma(\widehat Y[S])$, we consider an arbitrary subset $U\subseteq S$ with $|U|\geq 2$. Let $U_1,\ldots,U_c$ be the vertex sets of the connected components of $\widehat Y[U]$. Every $U_i$ is core-sparse because it is a subset of $S$. The same edge bound holds for singleton components, since they contain no edges. Applying the preceding bound to each component gives
    \[
    \begin{aligned}
        |E(\widehat Y[U])|
        &=\sum_{i=1}^{c}|E(\widehat Y[U_i])| \leq \left(1+\frac{\kappa_{\mathrm{ext}}}{\log n}\right) \sum_{i=1}^{c}(|U_i|-1)\\
        &=\left(1+\frac{\kappa_{\mathrm{ext}}}{\log n}\right)(|U|-c) \leq \left(1+\frac{\kappa_{\mathrm{ext}}}{\log n}\right)(|U|-1),
    \end{aligned}
    \]
    which further implies the third property.
\end{proof}

\subsection{Construction of Expander Towers}\label{subsec:expander-tower}

\begin{figure}[ht]
    \centering
    \includegraphics[width=0.9\linewidth]{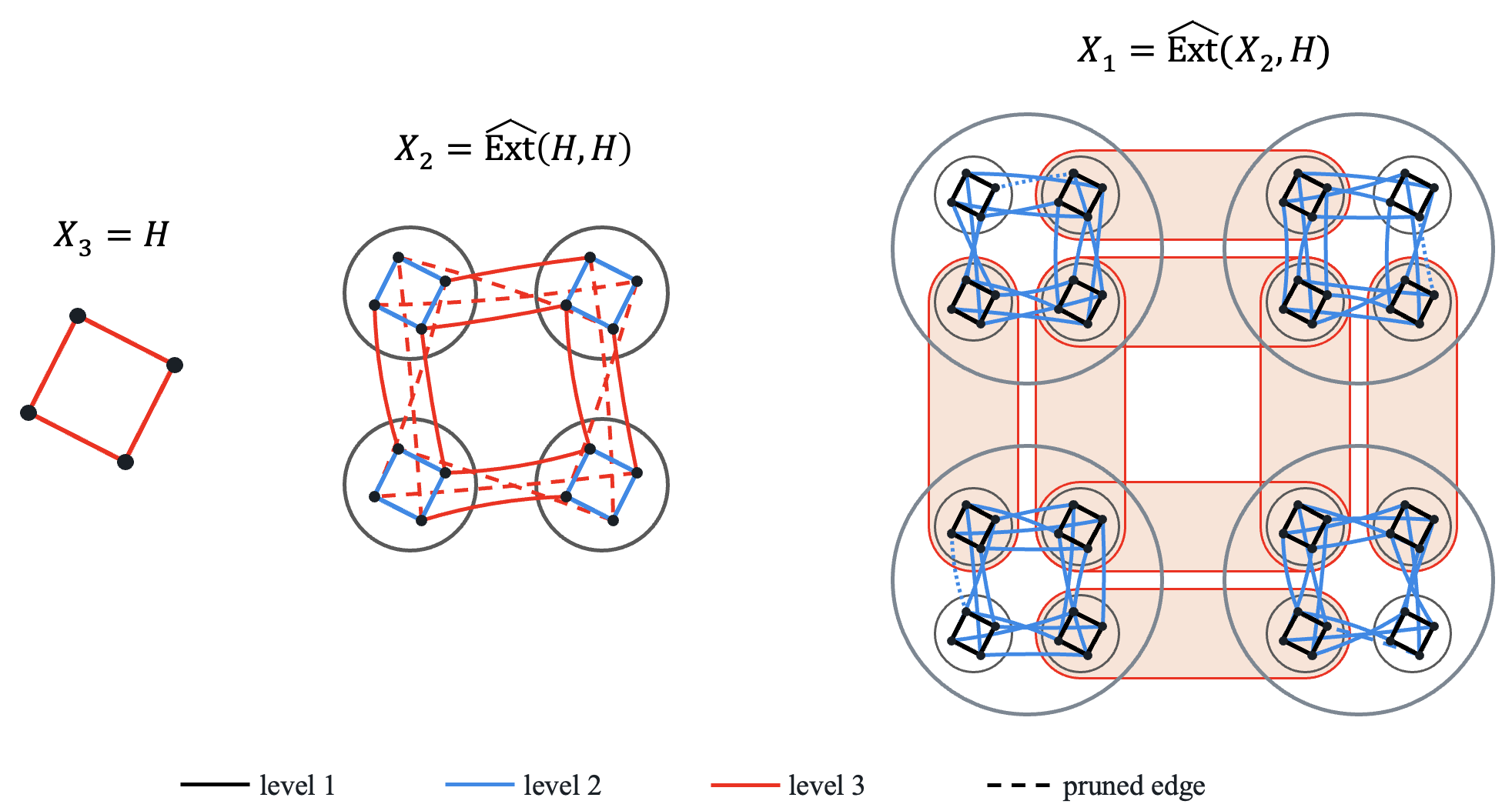}
    \caption{The figure shows a three-level expander tower. Black, blue, and red edges have levels $1$, $2$, and $3$, respectively. Dashed edges indicate deletions during pruning. In the rightmost graph, each shaded region represents a level-$3$ matching between the two enclosed cores, where some of the edges might be deleted in the pruning step.
\label{fig:expander-tower}
    }
    \label{figure:expander-tower}
\end{figure}

We are now ready to construct expander towers. Starting from $H$, we perform $t-1$ successive pruned extensions, each using $H$ as the core. We formalize this process as follows. See \Cref{figure:expander-tower} for an example with $t=3$.

\begin{definition}[Expander Tower]
    An \emph{expander tower} with base graph $H$ is a sequence of graphs $X_t, \ldots, X_1$ such that $X_t=H$ and for every $i= t-1, \ldots, 1$,
    \[
        X_i = \widehat{\Ext}(X_{i+1},H),
    \]
    where a realization of the pruned extension is chosen at each step. We define $X:=X_1$ as the \emph{final graph} of the tower.
\end{definition}

We observe that each extension adds one coordinate to the vertices of the contracted graph. For every $i\in[t]$, we therefore express every vertex of $X_i$ as a tuple $(x_t,\ldots,x_i)\in V(H)^{t-i+1}$. Since pruning deletes only edges, $|V(X_i)|=n^{t-i+1}$, and the final graph has $N:=|V(X)|=n^t$ vertices. For every $i\in [t-1]$, fixing the prefix $(x_t,\ldots,x_{i+1})$ identifies a core of $X_i$, whose vertices are indexed by the remaining coordinate $x_i\in V(H)$. We regard $X_t=H$ as its sole core and apply the definition of core-sparsity with respect to these cores in every $X_i$ for every $i\in[t]$. In particular, a set $S\subseteq V(X_t)$ is core-sparse precisely when $|S|\leq n^{0.9}$.

To relate vertices at different stages, we define the projections $p_i:V(X_i)\to V(X_{i+1})$ for every $i\in [t-1]$ and $\pi_i:V(X)\to V(X_i)$ for every $i\in[t]$ by
\[
\begin{aligned}
    p_i((x_t,\ldots,x_i))&:=(x_t,\ldots,x_{i+1}), \qquad \text{and}\\
    \pi_i((x_t,\ldots,x_1))&:=(x_t,\ldots,x_i).
\end{aligned}
\]
For every $i\in [t-1]$, the projection $p_i$ sends all vertices of a core to its corresponding contracted vertex and satisfies $p_i\circ\pi_i=\pi_{i+1}$. We emphasize that the map $\pi_1$ is the identity.

The same coordinates define a hierarchy within the final graph. For every $0\leq i\leq t$, an \emph{$i$-cloud} is the subset of $V(X)$ obtained by fixing the first $t-i$ coordinates $(x_t,\ldots,x_{i+1})$. It contains $n^i$ vertices. For every $i\in [t]$, fixing the additional coordinate $x_i$ partitions an $i$-cloud into $n$ $(i-1)$-clouds, called its \emph{children}. Thus, the $0$-clouds are singletons, the $1$-clouds are the vertex sets of the cores of $X$, and the unique $t$-cloud is the whole vertex set $V(X)$. For every $i\in[t]$, the projection $\pi_i$ maps an $i$-cloud onto the vertex set of a core in $X_i$, with each child mapping to one vertex of that core. This identifies the $n$ children of the cloud with the $n$ vertices of the corresponding core.

We next assign levels to edges during the construction. Every edge of $X_t=H$ has level $t$. For $i=t-1,\ldots,1$, the inter edges introduced in $X_i$ have level $i$, and each surviving outer edge inherits the level of its corresponding edge in $X_{i+1}$. For every $i\in[t]$ and every edge $\{u,v\}$ in a core of $X_i$, a \emph{level-$i$ lift} of $\{u,v\}$ is an edge $\{x,y\}\in E(X)$ satisfying
\[
    \{\pi_i(x),\pi_i(y)\}=\{u,v\}.
\]
Such an edge has level $i$ and joins the two children corresponding to $u$ and $v$ in the associated $i$-cloud.

For every $i\in[t]$, we assign each level-$i$ edge, in every graph of the tower, the length 
\[
    \ell_i:=(\log n)^{i/(t+1)}. 
\]
We further define $\sigma_i:=\ell_i\log n=(\log n)^{1+i/(t+1)}$. For every $i\in[t]$, let $d_{X_i}$ denotes the shortest-path distance with these edge lengths. Each core of $X_i$ is a copy of $H$ with all edges of length $\ell_i$. Hence, by \Cref{lemma:fa-expander}, every two vertices in the core can be joined within it by a path of length at most $\eta\sigma_i$. The following theorem shows that the realizations can be chosen so that all graphs in the tower are connected and the matching and fractional-arboricity guarantees of \Cref{lemma:pruned-extension} hold at every stage.

\begin{lemma}\label{lemma:expander-tower}
    There exists a constant $\kappa_{\mathrm{tow}}>0$ depending only on $t$ and an expander tower $X_t,\ldots,X_1$ with base graph $H$ satisfying:
    \begin{enumerate}
        \item for every $i\in[t]$, the graph $X_i$ is simple and connected, has $n^{t-i+1}$ vertices, maximum degree at most $10(t-i+1)$, and girth at least $g$;
        \item for every $i\in [t-1]$, each outer matching used to construct $X_i$ from $X_{i+1}$ retains at least $n-\lceil n^{0.99}\rceil$ edges after pruning;
        \item for every $i\in[t]$ and every core-sparse set $S\subseteq V(X_i)$,
        \[
            \Gamma(X_i[S])
            \leq 1+\frac{\kappa_{\mathrm{tow}}}{\log n}.
        \]
    \end{enumerate}
\end{lemma}

\begin{proof}
Let $\kappa$ and $\kappa_{\mathrm{ext}}$ be the constants in
\Cref{lemma:fa-expander,lemma:pruned-extension}, respectively, and set
\[
    \kappa_{\mathrm{tow}}
    :=\max\{\kappa,\kappa_{\mathrm{ext}}\}.
\]
This constant depends only on $t$. We construct the graphs in the order $X_t,\ldots,X_1$ and verify the properties inductively. The base graph $X_t=H$ is simple and connected, has $n$ vertices, maximum degree at most $3$, and girth at least $g$. Its connectedness follows from its positive edge expansion. Since a core-sparse subset of $V(X_t)$ has at most $n^{0.9}$ vertices, its fractional-arboricity bound follows from \Cref{lemma:fa-expander} and the choice of $\kappa_{\mathrm{tow}}$.

We now fix an integer $i\in [t-1]$, and suppose that $X_{i+1}$ has already been constructed with the required properties. By induction, $X_{i+1}$ is simple, has $|V(X_{i+1})|=n^{t-i}\leq n^{t-1}$ vertices, and has maximum degree at most $10(t-i)\leq 10(t-1)$. Thus, \Cref{lemma:pruned-extension} applies with $Q=X_{i+1}$. We choose a realization
\[
    X_i=\widehat{\Ext}(X_{i+1},H)
\]
given by that lemma. The resulting graph has girth at least $g$, every outer matching retains at least $n-\lceil n^{0.99}\rceil$ edges, and every core-sparse set $S\subseteq V(X_i)$ satisfies
\[
    \Gamma(X_i[S])
    \leq 1+\frac{\kappa_{\mathrm{ext}}}{\log n}
    \leq 1+\frac{\kappa_{\mathrm{tow}}}{\log n}.
\]

It remains to verify the other properties of $X_i$. The extension replaces each vertex of $X_{i+1}$ by a copy of $H$, so $X_i$ has $n^{t-i+1}$ vertices. It is simple because both $H$ and $X_{i+1}$ are simple. Each vertex has at most $3$ incident inter edges and at most $10(t-i)$ incident outer edges, giving maximum degree at most
\[
    3+10(t-i)\leq 10(t-i+1).
\]
Finally, every core is connected, and every outer matching contains a surviving edge because $n-\lceil n^{0.99}\rceil>0$ for sufficiently large $n$. The contracted graph $X_{i+1}$ is connected, so these surviving edges connect all the cores. Hence, $X_i$ is connected, completing the induction.
\end{proof}

We call an expander tower \emph{valid} if it satisfies the properties in \Cref{lemma:expander-tower}. The next section establishes structural properties that hold for every valid expander tower.

\subsection{Structural Properties}\label{subsec:tower-property}

In this section, we establish two properties of valid expander towers that will be used in the integrality gap analysis. We first compare distances at different stages of the construction, and then bound the number of surviving lifts between children of the same cloud.

\begin{lemma}\label{lemma:tower-distance}
    Let $X_t,\ldots,X_1$ be a valid expander tower with final graph $X=X_1$. For every $i\in [t-1]$ and $x,y\in V(X_i)$,
    \[
        d_{X_{i+1}}(p_i(x),p_i(y))
        \leq d_{X_i}(x,y)
        \leq d_{X_{i+1}}(p_i(x),p_i(y))+2\eta\sigma_i.
    \]
    Consequently, for every $r\in[t]$ and every $x,y\in V(X)$,
    \[
        d_{X_r}(\pi_r(x),\pi_r(y))
        \leq d_X(x,y)
        \leq d_{X_r}(\pi_r(x),\pi_r(y))
            +2\eta\sum_{i=1}^{r-1}\sigma_i.
    \]
    In particular, 
    \[
        \max_{x,y\in V(X)}d_X(x,y)\leq(2t-1)\eta\sigma_t.
    \]
\end{lemma}

\begin{proof}
For every $i\in [t-1]$, projecting a path in $X_i$ to $X_{i+1}$ contracts its inter edges and preserves the lengths of its outer edges. The resulting walk has no greater length, which proves the first inequality. To prove the upper bound, we will find a surviving lift of a shortest path in $X_{i+1}$. We first need a bound on the number of edges in that path. For every $i\in[t]$, let
\[
    \Lambda_i:=\max_{x,y\in V(X_i)}d_{X_i}(x,y).
\]
We prove the upper bounds in the order $i=t-1,\ldots,1$, while maintaining
$\Lambda_i\leq(2(t-i)+1)\eta\sigma_t$ for every $i\in[t]$. The initial bound $\Lambda_t\leq\eta\sigma_t$ follows from the unweighted diameter bound for $H$, since $X_t=H$ and its edges have length $\ell_t$.

We fix an integer $i\in [t-1]$ and assume that the diameter bound holds for $X_{i+1}$. We fix $x,y\in V(X_i)$ and let $P$ be a shortest path from $p_i(x)$ to $p_i(y)$ in $X_{i+1}$. Let $m:=|E(P)|$. Every edge of $P$ has length at least $\ell_1$, which implies that
\[
    m\leq\frac{\Lambda_{i+1}}{\ell_1}
    \leq 2t\eta\,\frac{\sigma_t}{\ell_1}
    \leq 2t\eta(\log n)^2,
\]
where the last inequality is because $\sigma_t/\ell_1=(\log n)^{2t/(t+1)}\leq(\log n)^2$. Since $t$ is fixed, this implies $m\lceil n^{0.99}\rceil<n$ for sufficiently large $n$.

We attempt to lift $P$ starting from each of the $n$ vertices in the core containing $x$. At each step, we follow the outer matching corresponding to the next edge of $P$, and we stop a lift if that matching edge was deleted. The matching property ensures that distinct starting vertices lead to distinct vertices after each step, as long as their lifts continue. Since at most $\lceil n^{0.99}\rceil$ edges were deleted from each matching, at most that many lifts stop at any given step. Thus, at most $m\lceil n^{0.99}\rceil<n$ starting vertices are excluded, and at least one lift survives in $X_i$. We choose such a lift and let $x',y'$ be its endpoints. The vertices $x,x'$ lie in the same core, as do $y,y'$. We join each pair within its core at a cost of at most $\eta\sigma_i$. The lift itself has the same length as $P$, because outer edges inherit the levels and lengths of the edges they replace. Therefore,
\[
    d_{X_i}(x,y)
    \leq d_{X_{i+1}}(p_i(x),p_i(y))+2\eta\sigma_i.
\]
This also gives
$\Lambda_i\leq\Lambda_{i+1}+2\eta\sigma_i
\leq(2(t-i)+1)\eta\sigma_t$,
since $\sigma_i\leq\sigma_t$. The induction is complete.

Finally, we fix $r\in[t]$ and $x,y\in V(X)$. For every $1\leq i<r$, we apply the distance comparison between $X_i$ and $X_{i+1}$ to the vertices $\pi_i(x)$ and $\pi_i(y)$. Since $p_i\circ\pi_i=\pi_{i+1}$, we obtain
\[
    0\leq
    d_{X_i}(\pi_i(x),\pi_i(y))
    -d_{X_{i+1}}(\pi_{i+1}(x),\pi_{i+1}(y))
    \leq 2\eta\sigma_i.
\]
We sum these inequalities over $i=1,\ldots,r-1$. The intermediate distance terms cancel, and $\pi_1$ is the identity, giving
\[
    0\leq
    d_X(x,y)-d_{X_r}(\pi_r(x),\pi_r(y))
    \leq 2\eta\sum_{i=1}^{r-1}\sigma_i.
\]
For $r=1$, the same conclusion holds directly because $\pi_1$ is the identity. The final weighted diameter bound follows from $\Lambda_1\leq(2t-1)\eta\sigma_t$, established by the induction above.
\end{proof}

We next show that adjacent vertices of a core give rise to a large matching between the corresponding child clouds. The proof follows one core edge through the remaining extensions and applies the matching guarantee at each step.

\begin{lemma}\label{lemma:cloud-matching}
    Let $X_t,\ldots,X_1$ be a valid expander tower with final graph $X=X_1$. Let $i\in[t]$, let $C$ be an $i$-cloud, and let $\{u,v\}$ be an edge in the core of $X_i$ corresponding to $C$. Let $C_u,C_v$ be the children of $C$ corresponding to $u,v$, respectively. Then, the level-$i$ lifts of $\{u,v\}$ in $X$ form a matching $M$ between $C_u$ and $C_v$ satisfying
    \[
        |M|\geq \bigl(n-\lceil n^{0.99}\rceil\bigr)^{i-1}
        \geq
        n^{i-1}\left(
            1-\frac{(t-1)\lceil n^{0.99}\rceil}{n}
        \right).
    \]
\end{lemma}

\begin{proof}
We follow $\{u,v\}$ through the $i-1$ extensions from $X_i$ to $X_1$. Initially, let $M_0:=\{\{u,v\}\}$. For every $0\leq r<i-1$, we obtain $M_{r+1}\subseteq E(X_{i-r-1})$ by taking all surviving edges in the outer matchings that replace the edges of $M_r\subseteq E(X_{i-r})$. We show inductively that $M_r$ is a matching and has at least $(n-\lceil n^{0.99}\rceil)^r$ edges for every $0\leq r\leq i-1$.

The claim holds for $M_0$, which consists of one edge. At an induction step, we fix an integer $r$ with $0\leq r<i-1$ and assume the claim holds for $M_r$. Each edge $\{x,y\}\in M_r$ is replaced by an outer matching between the new cores corresponding to $x$ and $y$. Since distinct edges of $M_r$ have disjoint endpoints, the cores associated with distinct edges are also disjoint. Hence, the surviving edges of these outer matchings together form a matching. By \Cref{lemma:expander-tower}, each retains at least $n-\lceil n^{0.99}\rceil$ edges, so
\[
    |M_{r+1}|
    \geq \bigl(n-\lceil n^{0.99}\rceil\bigr)|M_r|
    \geq \bigl(n-\lceil n^{0.99}\rceil\bigr)^{r+1}.
\]
This completes the induction. The final matching $M_{i-1}$ consists precisely of the level-$i$ lifts of $\{u,v\}$ in $X$, and its endpoints lie in $C_u$ and $C_v$. We obtain the second bound by applying 
\[
    (1-a)^{i-1}\geq 1-(i-1)a
\]
with $a=\lceil n^{0.99}\rceil/n$.
\end{proof}

\section{The Integrality Gap Instance}\label{sec:gap-instance}

In this section, we construct a $0$-Extension instance from a valid expander tower and show that its earthmover relaxation has a feasible solution whose cost is linear in the number of terminals. We define the instance in \Cref{subsec:gap-construction} and analyze the fractional solution in \Cref{subsec:fractional-solution}.

\subsection{Construction of the Instance}\label{subsec:gap-construction}

We fix an integer $t\geq 2$ and a sufficiently large even integer $n$. Let $X_t,\ldots,X_1$ be a valid expander tower as in \Cref{lemma:expander-tower}, with base graph $H$ and final graph $X:=X_1$. Let $N:=|V(X)|=n^t$. Recall that, for every $i\in[t]$, each level-$i$ edge has length $\ell_i=(\log n)^{i/(t+1)}$, and $\sigma_i=\ell_i\log n$. We define
\[
    \rho:=(\log n)^{(t-1)/(t+1)}
    \qquad\text{and}\qquad
    A:=\sigma_1=\ell_1\log n.
\]
In particular, $\rho=\ell_t/\ell_1=\sigma_t/A$.

We obtain $G=(V,E)$ from $X$ by adding, for every $x\in V(X)$, a distinct terminal $x_T$ and an edge $\{x,x_T\}$, which we call a \emph{terminal edge}. Thus,
\[
    T:=\{x_T:x\in V(X)\},\ \ \ \ 
    V:=V(X)\cup T,\ \ \ \  \text{and} \ \ \ \ 
    E:=E(X)\cup\bigl\{\{x,x_T\}:x\in V(X)\bigr\}.
\]
The graph $G$ has $2N$ vertices and $N$ terminals. We define the terminal metric using the shortest-path distances in $X$, i.e., for every $x,y\in V(X)$,
\[
    D(x_T,y_T):=d_X(x,y).
\]
We next assign weights $w:E\rightarrow \mathbb{R}_{\geq 0}$ to the edges of $G$ by setting
\[
    w_e:=
    \begin{cases}
        2\rho/\ell_1,
            & \text{if $e$ is a level-$1$ edge of $X$},\\
        1/\ell_i,
            & \text{if $e$ is a level-$i$ edge of $X$, where $2\leq i\leq t$},\\
        1/A,
            & \text{if $e$ is a terminal edge}.
    \end{cases}
\]
This defines the $0$-Extension instance $\mathcal{I}:=(G,w,T,D)$.

The edge lengths $\ell_i$ determine the terminal metric $D$, and the weights $w_e$ determine the objective value. For every $2\leq i\leq t$, the weight $1/\ell_i$ normalizes the cost of a level-$i$ edge by its length. The extra factor $2\rho$ on level-$1$ edges ensures that assigning different terminals to their endpoints costs at least $2\rho$, since distinct vertices of $X$ are at distance at least $\ell_1$. This penalty will be used in the integral lower bound in \Cref{sec:integral-lower-bound}. Each terminal edge charges the distance between a vertex's assigned terminal and its own terminal, divided by $A$.

\subsection{A Fractional Solution of Small Cost}\label{subsec:fractional-solution}

Our fractional solution assigns each vertex of $X$ the uniform distribution on the terminals corresponding to its $1$-cloud. These distributions agree within each $1$-cloud, so level-$1$ edges have zero cost. Each terminal edge costs at most $\eta$, by the diameter bound for its core. For edges between different $1$-clouds, the surviving outer matchings yield couplings of small cost.

\begin{lemma}\label{lemma:fractional-solution}
    The instance $\mathcal{I}$ constructed in \Cref{subsec:gap-construction} satisfies
    \[
        \operatorname{EM}(\mathcal{I})\leq (\eta+10t)N.
    \]
\end{lemma}

\begin{proof}
    For every $x\in V(X)$, let $F(x)$ be the $1$-cloud containing $x$, and let $\mu_x$ be the uniform distribution on $\{z_T:z\in F(x)\}$. For every terminal $x_T$, we choose $\mu_{x_T}$ with $\mu_{x_T}(x_T)=1$. These distributions form a feasible solution to the earthmover relaxation. If $\{x,y\}$ is a level-$1$ edge, then $F(x)=F(y)$ and $\mu_x=\mu_y$, so the edge contributes zero cost.

    For a terminal edge $\{x,x_T\}$, all mass in $\mu_x$ must be transported to $x_T$. The cloud $F(x)$ induces a copy of $H$ with edge lengths $\ell_1$. By \Cref{lemma:fa-expander}, every vertex of $F(x)$ can therefore reach $x$ within this core by a path of length at most $\eta\ell_1\log n=\eta A$. It follows that
    \[
        W_D(\mu_x,\mu_{x_T})
        =\frac{1}{n}\sum_{z\in F(x)}d_X(z,x)
        \leq\eta A.
    \]
    Multiplying by the weight $1/A$, we obtain a cost of at most $\eta$ for each terminal edge.

    We now fix an integer $2\leq i\leq t$ and a level-$i$ edge $\{x,y\}\in E(X)$. In the final extension $X=\widehat{\Ext}(X_2,H)$, this edge belongs to a surviving outer matching $M$ between $F(x)$ and $F(y)$. Every edge of $M$ has level $i$, and \Cref{lemma:expander-tower} gives $|M|\geq n-\lceil n^{0.99}\rceil$. We couple $\mu_x$ and $\mu_y$ as follows. For every edge $\{u,v\}\in M$, with $u\in F(x)$ and $v\in F(y)$, we send mass $1/n$ from $u_T$ to $v_T$. Each such pair has distance $D(u_T,v_T)=d_X(u,v)\leq\ell_i$. Each cloud has exactly $n-|M|$ unmatched vertices. We pair these vertices by an arbitrary bijection and send mass $1/n$ between the corresponding terminals. Every terminal in either support appears in exactly one pair, so the resulting transportation plan is a coupling of $\mu_x$ and $\mu_y$.

    By \Cref{lemma:tower-distance}, the diameter of $X$ is at most $(2t-1)\eta\sigma_t$, which bounds the distance of every unmatched pair. This coupling gives
    \[
    \begin{aligned}
        W_D(\mu_x,\mu_y)
        &\leq \frac{|M|}{n}\ell_i
            +\frac{n-|M|}{n}(2t-1)\eta\sigma_t\\
        &\leq \ell_i+(2t-1)\eta
            \frac{\lceil n^{0.99}\rceil}{n}\sigma_t.
    \end{aligned}
    \]
    Since $\{x,y\}$ has weight $1/\ell_i$, its contribution is at most
    \[
        w_{\{x,y\}}W_D(\mu_x,\mu_y)
        \leq 1+(2t-1)\eta
            \frac{\lceil n^{0.99}\rceil}{n}
            \frac{\sigma_t}{\ell_i}
        =1+o(1),
    \]
    where the last equality follows from
    $\lceil n^{0.99}\rceil/n=O(n^{-0.01})$ and
    $\sigma_t/\ell_i=(\log n)^{1+(t-i)/(t+1)}$.
    Thus, for sufficiently large $n$, every edge of level at least $2$ has cost at most $2$.

    We recall that there are $N$ terminal edges, each of cost at most $\eta$. Every edge of $X$ has cost at most $2$, and \Cref{lemma:expander-tower} gives $|E(X)|\leq 5tN$. Hence, the total cost is at most
    \[
        \eta N+2|E(X)|\leq(\eta+10t)N.
    \]
\end{proof}

To establish an $\Omega(\rho)$ integrality gap, it remains to show that every integral assignment has cost at least $c_tN\rho$ for a constant $c_t>0$ depending only on $t$. We develop the required random partition lemma in \Cref{sec:random-partitions} and prove the integral lower bound in \Cref{sec:integral-lower-bound}.

\section{Random Partitions of Small Vertex Sets}\label{sec:random-partitions}

In this section, we construct random partitions of small vertex sets in an expander tower. The parts have small diameter, and nearby vertices are unlikely to lie in different parts. In Section~\ref{sec:integral-lower-bound}, we apply these partitions to the representatives of the children of a cloud to choose a representative for their parent. Throughout this section, we fix an integer $t\geq 2$ and set $\tau_0:=1/(200\log(20t))$.

For a partition $\mathcal P$ of a set $Z$, we let $\mathcal P(x)$ denote the part containing $x$ for every $x\in Z$. Let $G=(V,E)$ be a graph with positive edge lengths. A partition of $Z\subseteq V$ is \emph{$\Lambda$-bounded} if every two vertices in the same part have distance at most $\Lambda$ in $G$; a random partition is $\Lambda$-bounded if this holds with probability one. Distances are measured in the whole graph $X$, so shortest paths may leave $Z$. The same terminology applies to finite metric spaces. We also let $B_G(x,r):=\{y\in V:d_G(x,y)\leq r\}$ for every $x\in V$ and $r\geq 0$, and discard empty parts from all partitions of vertex sets.

\begin{lemma}\label{lemma:tower-partition}
    There is a constant $K_t\geq 1$ with the following property. For every fixed $0<\tau\leq\tau_0$ and every sufficiently large even integer $n$, let $X_t,\ldots,X_1$ be a valid expander tower with an $n$-vertex base graph and final graph $X:=X_1$. For every $Z\subseteq V(X)$ with $|Z|\leq n$ and every $2\leq i\leq t$, there exists a $\tau\sigma_i$-bounded random partition $\mathcal P$ of $Z$, with respect to $d_X$, satisfying
    \[
        \Pr[\mathcal P(x)\neq\mathcal P(y)]
        \leq K_t\frac{d_X(x,y)}{\sigma_{i-1}}
        \qquad\text{for every }x,y\in Z.
    \]
\end{lemma}

The constant $K_t$ depends only on $t$, whereas the required lower bound on $n$ may also depend on $\tau$. We first construct partitions from small fractional arboricity by sampling forests in Section~\ref{subsec:fa-partitions}, and show how to extend a partition through a forest in Section~\ref{subsec:forest-extension}. We combine these ingredients for two consecutive tower levels in Section~\ref{subsec:adjacent-level-partitions}, then transfer the resulting partition to the final graph in Section~\ref{subsec:final-graph-partitions}.

\subsection{Partitions from Small Fractional Arboricity}\label{subsec:fa-partitions}

We first show that a graph with fractional arboricity close to one admits a distribution over forests that retains each edge with high probability. We use the following theorem of Nash--Williams, which characterizes when the edges of a graph can be partitioned into forests.

\begin{theorem}[Nash--Williams~\cite{nash1964decomposition}]\label{theorem:forest-decomposition}
    Let $G=(V,E)$ be a multigraph without self-loops, and let $a\geq 1$ be an integer. The edge set $E$ can be partitioned into $a$ forests if and only if
    \[
        |E(G[S])|\leq a(|S|-1)
        \qquad\text{for every }S\subseteq V\text{ with }|S|\geq 2.
    \]
\end{theorem}

To obtain a forest distribution, we apply this theorem after replacing each edge by the same number of parallel copies, and then choose one forest uniformly at random.

\begin{lemma}\label{lemma:forest-sampling}
    Let $\theta\geq 0$ and $G=(V,E)$ be a graph without self-loops satisfying $\Gamma(G)\leq 1+\theta$. There exists a distribution over spanning forests $F$ of $G$ such that
    \[
        \Pr[e\in E(F)]\geq\frac{1}{1+\theta}
        \qquad\text{for every }e\in E.
    \]
    In particular, $\Pr[e\notin E(F)]\leq\theta$ for every $e\in E$.
\end{lemma}

\begin{proof}
    Since $\Gamma(G)$ is rational, we may choose positive integers $a,b$ such that $\Gamma(G)=a/b$. We replace every edge of $G$ by $b$ parallel copies to obtain a multigraph $G'=(V,E')$. For every $S\subseteq V$ with $|S|\geq 2$,
    \[
        |E(G'[S])|
        =b|E(G[S])|
        \leq b\Gamma(G)(|S|-1)
        =a(|S|-1).
    \]
    By \Cref{theorem:forest-decomposition}, $E'$ can therefore be partitioned into $a$ forests. We choose one uniformly at random and replace each selected copy by its original edge in $G$. The $b$ copies of any edge belong to distinct forests, since two parallel edges form a cycle. Thus, the resulting forest contains each edge with probability $b/a=1/\Gamma(G)\geq 1/(1+\theta)$. We include all vertices of $V$ and add edges, without creating cycles, until the forest contains a spanning tree of every connected component of $G$. This only increases edge inclusion probabilities, and $1-1/(1+\theta)\leq\theta$ gives the final assertion.
\end{proof}

We next partition each sampled tree by cutting it at randomly shifted distances from its root. An edge can have its endpoints separated either because it is omitted from the forest or because it is cut in this second step. These two events account for the two terms in the following bound.

\begin{lemma}\label{lemma:fa-partition}
    Let $\theta\geq 0$, and let $G=(V,E)$ be a graph without self-loops satisfying $\Gamma(G)\leq 1+\theta$, with edge lengths at least $a>0$. For every $\Lambda>0$, there exists a $\Lambda$-bounded random partition $\mathcal P$ of $V$ such that
    \[
        \Pr[\mathcal P(x)\neq\mathcal P(y)]
        \leq
        \left(\frac{2}{\Lambda}+\frac{\theta}{a}\right)d_G(x,y)
    \]
    for every $x,y\in V$ that are connected in $G$.
\end{lemma}

\begin{proof}
    We sample a spanning forest $F$ using \Cref{lemma:forest-sampling}. For every tree $T=(V_T,E_T)$ of $F$, we choose a root and let $h(v)$ be the distance from the root to $v\in V_T$ within $T$. We independently choose $U_T$ uniformly from $[0,\Lambda/2)$ and assign to every $v\in V_T$ the integer
    $b(v):=\lfloor 2(h(v)+U_T)/\Lambda\rfloor$.
    We delete every edge $\{u,v\}$ of $T$ with $b(u)\neq b(v)$ and take the connected components of the resulting forest as the parts of $\mathcal P$. To verify the diameter bound, let $x,y$ belong to the same part, and let $z$ be their least common ancestor. The path from $x$ to $y$ remains in their part, so $x,y,z$ have the same value of $b$. Both $h(x)-h(z)$ and $h(y)-h(z)$ are therefore less than $\Lambda/2$, giving
    \[
        d_G(x,y)
        \leq d_T(x,y)
        =h(x)-h(z)+h(y)-h(z)
        <\Lambda.
    \]

    We next bound the separation probability. Let $\ell_e$ be the length of an edge $e=\{u,v\}\in E$. Conditional on a forest containing $e$, the depths of its endpoints differ by $\ell_e$, so the random shift separates them with probability $\min\{1,2\ell_e/\Lambda\}$. Since $e$ is omitted from the forest with probability at most $\theta$, its endpoints are separated with probability at most $\theta+2\ell_e/\Lambda$. For connected vertices $x,y\in V$, let $P$ be a shortest path between them. If $x$ and $y$ lie in different parts, some edge of $P$ has its endpoints in different parts. As $P$ contains at most $d_G(x,y)/a$ edges, a union bound gives
    \[
        \Pr[\mathcal P(x)\neq\mathcal P(y)]
        \leq
        \sum_{e\in E(P)}
        \left(\theta+\frac{2\ell_e}{\Lambda}\right)
        \leq
        \left(\frac{\theta}{a}+\frac{2}{\Lambda}\right)d_G(x,y).
    \]
\end{proof}

\subsection{Extending Partitions through Forests}\label{subsec:forest-extension}

We next show how to extend a partition of a selected set $A$ through a forest. In each tree meeting $A$, we assign vertices to nearby vertices of $A$ so that they can inherit parts of the given partition. We use a random retraction for this assignment. For a finite metric space $(M,d)$ and a nonempty set $B\subseteq M$, let $d(x,B):=\min_{b\in B}d(x,b)$ for every $x\in M$. A map $g:M\to B$ is a \emph{retraction} if $g(b)=b$ for every $b\in B$. The following theorem constructs a random retraction from random partitions of $(M,d)$.

\begin{theorem}[{\cite{lee2009geometry}}]\label{theorem:random-retraction}
    Let $(M,d)$ be a finite metric space, let $B\subseteq M$ be nonempty, and let $\beta\geq 0$. Suppose that for every $\Lambda>0$ there is a $\Lambda$-bounded random partition $\mathcal Q_\Lambda$ of $M$ satisfying
    \[
        \Pr[\mathcal Q_\Lambda(x)\neq\mathcal Q_\Lambda(y)]
        \leq\beta\frac{d(x,y)}{\Lambda}
        \qquad\text{for every }x,y\in M.
    \]
    Then, there exists a random retraction $g:M\to B$ such that $d(x,g(x))\leq 5d(x,B)$ for every $x\in M$. Moreover, for every $x,y\in M\setminus B$,
    \[
        \Pr[g(x)\neq g(y)]
        \leq(\beta+2)\frac{d(x,y)}{\min\{d(x,B),d(y,B)\}}.
    \]
\end{theorem}

For a tree metric, \Cref{lemma:fa-partition} supplies the required partitions with $\beta=2$. Combining the two guarantees in \Cref{theorem:random-retraction} gives the following bound on the expected distance between the images of two vertices.

\begin{lemma}\label{lemma:tree-retraction}
    There is an absolute constant $C_0\geq 1$ with the following property. Let $T=(V_T,E_T)$ be a tree with positive edge lengths, and let $B\subseteq V_T$ be nonempty. There is a random retraction $g:V_T\to B$ such that $d_T(v,g(v))\leq 5d_T(v,B)$ for every $v\in V_T$, and
    \[
        \mathbb E[d_T(g(u),g(v))]\leq C_0d_T(u,v)
        \qquad\text{for every }u,v\in V_T.
    \]
\end{lemma}

\begin{proof}
    By \Cref{lemma:fa-partition} with $\theta=0$, the metric $(V_T,d_T)$ satisfies the hypothesis of \Cref{theorem:random-retraction} with $\beta=2$. We obtain a random retraction $g$ with the required bound on $d_T(v,g(v))$. To bound the expected distance between its images, we fix $u,v\in V_T$ and set $d:=d_T(u,v)$ and $q:=\min\{d_T(u,B),d_T(v,B)\}$. The triangle inequality gives $\max\{d_T(u,B),d_T(v,B)\}\leq q+d$, and hence,
    \[
        d_T(g(u),g(v))
        \leq 5d_T(u,B)+d+5d_T(v,B)
        \leq 10q+6d.
    \]
    If $q\leq d$, this bounds the expectation by $16d$. If $q>d$, the displayed bound is at most $16q$, and \Cref{theorem:random-retraction} gives $\Pr[g(u)\neq g(v)]\leq 4d/q$. Since the distance between the images is zero whenever they coincide, in this case
    \[
        \mathbb E[d_T(g(u),g(v))]
        \leq 16q\Pr[g(u)\neq g(v)]
        \leq 64d.
    \]
    Thus, $C_0=64$ suffices.
\end{proof}

We now extend a partition of a set $A$ to a graph whose edges with at least one endpoint outside $A$ form a forest. Vertices close to $A$ in this forest inherit the part of their image under the retraction, while the remaining vertices use a partition of the forest itself. We choose the distance threshold at random so that nearby vertices are unlikely to fall on opposite sides of it.

\begin{lemma}\label{lemma:forest-extension}
    There is an absolute constant $c>0$ with the following property. Let $G=(V,E)$ be a graph with positive edge lengths, and let $A\subseteq V$ such that $F:=(V,E\setminus E(G[A]))$ is a forest. Let $\Lambda>0$ and $\beta\geq 0$. Suppose that $A$ has a $\Lambda/2$-bounded random partition $\mathcal P_A$, with respect to $d_G$, satisfying
    \[
        \Pr[\mathcal P_A(u)\neq\mathcal P_A(v)]
        \leq\beta\frac{d_G(u,v)}{\Lambda}
        \qquad\text{for every }u,v\in A\text{ connected in }G.
    \]
    Then, there exists a $\Lambda$-bounded random partition $\mathcal P$ of $V$ whose restriction to $A$ is $\mathcal P_A$ and such that
    \[
        \Pr[\mathcal P(x)\neq\mathcal P(y)]
        \leq c\max\{8,\beta\}\frac{d_G(x,y)}{\Lambda}
    \]
    for every $x,y\in V$ that are connected in $G$.
\end{lemma}

\begin{proof}
    We sample $\mathcal P_A$ from the given distribution and choose $r$ uniformly from $[\Lambda/40,\Lambda/20]$. For every tree $T=(V_T,E_T)$ of $F$, we set $B_T:=A\cap V_T$ and sample a $\Lambda$-bounded partition $\mathcal Q_T$ using \Cref{lemma:fa-partition} with $\theta=0$. If $B_T\neq\varnothing$, we also sample a retraction $g_T:V_T\to B_T$ using \Cref{lemma:tree-retraction}. All these random choices are independent. On every tree meeting $A$, we define $q(v):=d_T(v,B_T)$ and $g(v):=g_T(v)$ for every $v\in V_T$. We then extend each part $C\in\mathcal P_A$ to
    \[
        C^+:=
        \bigcup_{\substack{T\text{ a tree of }F\\B_T\neq\varnothing}}
        \{v\in V_T:q(v)\leq r,\ g(v)\in C\}.
    \]
    Within every tree meeting $A$, we partition the remaining vertices into the sets
    $\{v\in D:q(v)>r\}$ for every $D\in\mathcal Q_T$.
    For every tree disjoint from $A$, we retain all parts of $\mathcal Q_T$. Together, these sets form $\mathcal P$. Since $q(a)=0$ and $g(a)=a$ for every $a\in A$, the restriction of $\mathcal P$ to $A$ is exactly $\mathcal P_A$.

    We first prove the diameter bound. Let $x,y$ belong to the same part of $\mathcal P$. If this part is $C^+$ for some $C\in\mathcal P_A$, then $g(x),g(y)\in C$. Each vertex is within distance $5r$ of its image, so
    \[
    \begin{aligned}
        d_G(x,y)
        &\leq d_G(x,g(x))+d_G(g(x),g(y))+d_G(g(y),y)\\
        &\leq 5q(x)+\Lambda/2+5q(y)\\
        &\leq 10r+\Lambda/2
        \leq\Lambda.
    \end{aligned}
    \]
    Otherwise, their part is contained in a part of $\mathcal Q_T$ for some tree $T$, and $d_G(x,y)\leq d_T(x,y)\leq\Lambda$.

    We next bound the separation probability for an edge $e=\{u,v\}$ of a tree $T$ meeting $A$, and let $\ell_e$ be its length. If its endpoints lie in different parts, then exactly one of $q(u),q(v)$ is at most $r$, or their images lie in different parts of $\mathcal P_A$, or they lie in different parts of $\mathcal Q_T$. Since $|q(u)-q(v)|\leq\ell_e$ and $r$ is uniform on an interval of length $\Lambda/40$, the first event has probability at most $40\ell_e/\Lambda$. For the second event, independence of $g$ and $\mathcal P_A$ and \Cref{lemma:tree-retraction} give
    \[
    \begin{aligned}
        \Pr[\mathcal P_A(g(u))\neq\mathcal P_A(g(v))]
        &\leq\frac{\beta}{\Lambda}
            \mathbb E[d_G(g(u),g(v))]\\
        &\leq\frac{\beta}{\Lambda}
            \mathbb E[d_T(g(u),g(v))]\\
        &\leq\frac{\beta C_0\ell_e}{\Lambda}.
    \end{aligned}
    \]
    The third event has probability at most $2\ell_e/\Lambda$ by \Cref{lemma:fa-partition}, so the endpoints of $e$ are separated with probability at most $(42+C_0\beta)\ell_e/\Lambda$. This bound also holds for every other edge of $G$: an edge in $G[A]$ has separation probability at most $\beta\ell_e/\Lambda$, since the partition on $A$ is unchanged, and an edge in a tree disjoint from $A$ has separation probability at most $2\ell_e/\Lambda$. A union bound along a shortest path between connected vertices $x,y\in V$ now gives
    \[
        \Pr[\mathcal P(x)\neq\mathcal P(y)]
        \leq(42+C_0\beta)\frac{d_G(x,y)}{\Lambda}
        \leq c\max\{8,\beta\}\frac{d_G(x,y)}{\Lambda},
    \]
    where we may take $c:=C_0+6$.
\end{proof}

\subsection{Partitions across Consecutive Tower Levels}\label{subsec:adjacent-level-partitions}

We now combine the preceding lemmas for two consecutive tower levels. A neighborhood of the set to be partitioned may contain too many vertices in a single core for the fractional-arboricity bound to apply directly. We partition these cores in the contracted graph, then use a sampled forest to extend the partition to the remaining vertices. Throughout this section, we fix $0<\tau\leq\tau_0$ and a sufficiently large even integer $n$, and let $X_t,\ldots,X_1$ be a valid expander tower with an $n$-vertex base graph. We retain its fractional-arboricity constant $\kappa_{\mathrm{tow}}$ from \Cref{lemma:expander-tower}.

\begin{lemma}\label{lemma:adjacent-level-partition}
    There is a constant $K_t\geq 1$, independent of $\tau$ and $n$, such that, for every $2\leq i\leq t$ and every $Z\subseteq V(X_{i-1})$ with $|Z|\leq n$, there exists a $\tau\sigma_i$-bounded random partition $\mathcal P$ of $Z$, with respect to $d_{X_{i-1}}$, satisfying
    \[
        \Pr[\mathcal P(x)\neq\mathcal P(y)]
        \leq K_t\frac{d_{X_{i-1}}(x,y)}{\sigma_{i-1}}
        \qquad\text{for every }x,y\in Z.
    \]
\end{lemma}

We fix an integer $2\leq i\leq t$ and a set $Z\subseteq V(X_{i-1})$ with $|Z|\leq n$. Let $Y:=X_{i-1}$, $Q:=X_i$, and let $p:V(Y)\to V(Q)$ be the projection $p_{i-1}$. For every $u\in V(Q)$, let $H_u$ be the corresponding core of $Y$. We set $\Lambda:=\tau\sigma_i$ and define
\[
    U:=\bigcup_{z\in Z}B_Y(z,\tau_0\sigma_{i-1}).
\]
We call a core \emph{heavy} if it contains more than $n^{0.75}$ vertices of $U$. Let
\[
    R:=\{u\in V(Q):|U\cap V(H_u)|>n^{0.75}\},
    \qquad
    A:=\bigcup_{u\in R}V(H_u).
\]
Thus, $R$ indexes the heavy cores and $A$ contains all their vertices, including those outside $U$. We define $G=(V,E)$ by $V:=U\cup A$ and $E:=E(Y[V])\setminus E(Y[A])$, so $G$ retains precisely the edges with at least one endpoint outside $A$. The next lemma bounds the number of heavy cores and shows that $G$ has fractional arboricity close to one.

\begin{lemma}\label{lemma:heavy-core-sparsity}
    The sets $U$ and $R$ satisfy $|U|\leq n^{1.01}$ and $|R|\leq n^{0.26}$. Moreover, the set of vertices incident to an edge of $G$ is core-sparse in $Y$, and
    \[
        \Gamma(G)\leq 1+\frac{\kappa_{\mathrm{tow}}}{\log n}.
    \]
\end{lemma}

\begin{proof}
    Every edge of $Y$ has length at least $\ell_{i-1}$. Thus, every vertex in $B_Y(z,\tau_0\sigma_{i-1})$ can be reached from $z$ by a path containing at most $\tau_0\log n$ edges. Since $Y$ has maximum degree at most $10t$,
    \[
        |U|
        \leq 2n(10t)^{1+\tau_0\log n}
        \leq n^{1.01}
    \]
    for sufficiently large $n$. The heavy cores are disjoint and each contains more than $n^{0.75}$ vertices of $U$, so
    $|R|\leq |U|/n^{0.75}\leq n^{0.26}$.

    Let $S$ be the set of vertices incident to an edge of $G$. A core that is not heavy contains at most $n^{0.75}$ vertices of $V$, and hence at most that many vertices of $S$. A vertex of $S$ in a heavy core must be joined by an edge of $G$ to a core that is not heavy, since $G$ contains no edge with both endpoints in $A$. Each core has at most $10t$ neighboring cores, and the edges between any two cores form a matching. Each neighboring core that is not heavy therefore accounts for at most $n^{0.75}$ vertices of $S$ in the heavy core. Thus, every core contains at most $10t n^{0.75}\leq n^{0.9}$ vertices of $S$ for sufficiently large $n$. The set $S$ is consequently core-sparse, and validity of the tower gives $\Gamma(Y[S])\leq 1+\kappa_{\mathrm{tow}}/\log n$. Since $G$ is obtained from a subgraph of $Y[S]$ by adding isolated vertices, it satisfies the same fractional-arboricity bound.
\end{proof}

To partition $A$, we partition the corresponding vertices $R$ in $Q$ and place each entire core in the part of its contracted vertex. A neighborhood of $R$ is small enough to be core-sparse, so \Cref{lemma:fa-partition} applies. We then use \Cref{lemma:tower-distance} to bound the additional diameter when contracted vertices are replaced by their cores.

\begin{lemma}\label{lemma:heavy-core-partition}
    The set $A$ has a $\Lambda/2$-bounded random partition $\mathcal P_A$, with respect to $d_Y$, satisfying
    \[
        \Pr[\mathcal P_A(x)\neq\mathcal P_A(y)]
        \leq
        (8+\kappa_{\mathrm{tow}}\tau)
        \frac{d_Y(x,y)}{\Lambda}
        \qquad\text{for every }x,y\in A.
    \]
\end{lemma}

\begin{proof}
    If $A=\varnothing$, the empty partition satisfies the claim. Otherwise, we consider $W:=\bigcup_{u\in R}B_Q(u,\Lambda)$. Every edge of $Q$ has length at least $\ell_i$, and $\Lambda/\ell_i=\tau\log n$. By the degree bound for $Q$ and \Cref{lemma:heavy-core-sparsity},
    \[
        |W|
        \leq 2n^{0.26}(10t)^{1+\tau\log n}
        \leq n^{0.27}
        <n^{0.9}
    \]
    for sufficiently large $n$, where we use $\tau\leq\tau_0$. Hence, $W$ is core-sparse in $Q$ and $\Gamma(Q[W])\leq 1+\kappa_{\mathrm{tow}}/\log n$. We apply \Cref{lemma:fa-partition} to $Q[W]$ with diameter parameter $\Lambda/4$ and minimum edge length $\ell_i$, obtaining a random partition $\mathcal Q$ of $W$. For every $u,v\in R$ with $d_Q(u,v)\leq\Lambda$, a shortest path between them lies in $B_Q(u,\Lambda)\subseteq W$. Thus, $d_{Q[W]}(u,v)=d_Q(u,v)$, and
    \[
    \begin{aligned}
        \Pr[\mathcal Q(u)\neq\mathcal Q(v)]
        &\leq
        \left(\frac{8}{\Lambda}
        +\frac{\kappa_{\mathrm{tow}}}{\ell_i\log n}\right)d_Q(u,v)\\
        &=(8+\kappa_{\mathrm{tow}}\tau)
            \frac{d_Q(u,v)}{\Lambda}.
    \end{aligned}
    \]
    If $d_Q(u,v)>\Lambda$, the same bound follows from the probability being at most one. Thus, it holds for every $u,v\in R$.

    We construct $\mathcal P_A$ by replacing each part $C\in\mathcal Q$ with $\{x\in A:p(x)\in C\}$. If $x,y$ lie in the same resulting part, their projections lie in the same part of $\mathcal Q$. By \Cref{lemma:tower-distance},
    \[
        d_Y(x,y)
        \leq d_Q(p(x),p(y))+2\eta\sigma_{i-1}
        \leq\Lambda/4+2\eta\sigma_{i-1}
        \leq\Lambda/2.
    \]
    The last inequality holds for sufficiently large $n$, since
    $\Lambda/\sigma_{i-1}=\tau(\log n)^{1/(t+1)}$.
    The separation bound for $\mathcal Q$ and the inequality $d_Q(p(x),p(y))\leq d_Y(x,y)$ give, for every $x,y\in A$,
    \[
    \begin{aligned}
        \Pr[\mathcal P_A(x)\neq\mathcal P_A(y)]
        &=\Pr[\mathcal Q(p(x))\neq\mathcal Q(p(y))]\\
        &\leq(8+\kappa_{\mathrm{tow}}\tau)
            \frac{d_Y(x,y)}{\Lambda}.
    \end{aligned}
    \]
\end{proof}

We can now complete the construction. We sample a forest of $G$ and add edges between vertices of $A$ whose lengths equal their distances in $Y$. These added edges preserve the metric on $A$, so \Cref{lemma:forest-extension} applies to the partition just constructed. We then restrict the extended partition to $Z$.

\begin{proof}[Proof of \Cref{lemma:adjacent-level-partition}]
    By \Cref{lemma:heavy-core-sparsity,lemma:forest-sampling}, we sample a spanning forest $F$ of $G$ with $\Pr[e\notin E(F)]\leq\kappa_{\mathrm{tow}}/\log n$ for every $e\in E$. We obtain a graph $G_F$ on $V$ by adding to $F$, for every pair of distinct vertices $u,v\in A$, an edge of length $d_Y(u,v)$. Removing the edges with both endpoints in $A$ leaves exactly $F$. Every edge of $G_F$ has length at least the distance in $Y$ between its endpoints, so the triangle inequality gives
    $d_Y(u,v)\leq d_{G_F}(u,v)$
    whenever $u,v$ are connected in $G_F$. For every $u,v\in A$, the added edge also gives the reverse inequality, so
    \[
        d_{G_F}(u,v)=d_Y(u,v)
        \qquad\text{for every }u,v\in A.
    \]

    Independently of $F$, we sample $\mathcal P_A$ using \Cref{lemma:heavy-core-partition}. Since $d_{G_F}$ and $d_Y$ agree on $A$, this partition satisfies the hypotheses of \Cref{lemma:forest-extension} for $G_F$, with $\beta:=8+\kappa_{\mathrm{tow}}\tau$. Conditional on $F$, that lemma gives a $\Lambda$-bounded random partition $\mathcal R$ of $V$, with respect to $d_{G_F}$, satisfying
    \begin{equation}\label{eq:conditional-forest-partition}
        \Pr[\mathcal R(u)\neq\mathcal R(v)\mid F]
        \leq B_t\frac{d_{G_F}(u,v)}{\Lambda}
    \end{equation}
    for every $u,v\in V$ connected in $G_F$, where $B_t:=c(8+\kappa_{\mathrm{tow}}\tau_0)$ depends only on $t$. We restrict $\mathcal R$ to $Z$ to obtain
    $\mathcal P:=\{C\cap Z:C\in\mathcal R\}$.
    Since $d_Y\leq d_{G_F}$ on connected pairs, every part of $\mathcal P$ has diameter at most $\Lambda=\tau\sigma_i$ in $Y$.

    We next bound separation probabilities along the edges of $Y[V]$. Let $e=\{u,v\}$ be such an edge, and let $\ell_e$ be its length. If at least one endpoint lies outside $A$, then $e\in E$. The edge is omitted from $F$ with probability at most $\kappa_{\mathrm{tow}}/\log n$. Whenever it is retained, $d_{G_F}(u,v)\leq\ell_e$, so \eqref{eq:conditional-forest-partition} bounds its separation probability by $B_t\ell_e/\Lambda$. Averaging over $F$ gives
    \[
        \Pr[\mathcal R(u)\neq\mathcal R(v)]
        \leq\frac{\kappa_{\mathrm{tow}}}{\log n}
            +B_t\frac{\ell_e}{\Lambda}.
    \]
    If both endpoints lie in $A$, then
    $d_{G_F}(u,v)=d_Y(u,v)\leq\ell_e$
    for every $F$, so the same bound holds without the first term. Thus, the displayed bound applies to every edge of $Y[V]$.

    Finally, we set $K_t:=\max\{B_t+\kappa_{\mathrm{tow}},\tau_0^{-1},1\}$ and fix $x,y\in Z$. If $d_Y(x,y)>\tau_0\sigma_{i-1}$, then
    $K_t d_Y(x,y)/\sigma_{i-1}\geq 1$,
    so the desired separation bound is immediate. Otherwise, a shortest path $P$ from $x$ to $y$ lies in
    $B_Y(x,\tau_0\sigma_{i-1})\subseteq U$.
    Every edge of $P$ has length at least $\ell_{i-1}$, so
    $|E(P)|\leq d_Y(x,y)/\ell_{i-1}$.
    Since $\mathcal P$ is the restriction of $\mathcal R$, summing the edge separation bounds along $P$ gives
    \[
    \begin{aligned}
        \Pr[\mathcal P(x)\neq\mathcal P(y)]
        &\leq
        \left(
            \frac{\kappa_{\mathrm{tow}}}{\ell_{i-1}\log n}
            +\frac{B_t}{\tau\sigma_i}
        \right)d_Y(x,y)\\
        &=
        \left(
            \kappa_{\mathrm{tow}}
            +\frac{B_t}{\tau(\log n)^{1/(t+1)}}
        \right)
        \frac{d_Y(x,y)}{\sigma_{i-1}}.
    \end{aligned}
    \]
    For fixed $\tau$ and sufficiently large $n$, the coefficient in parentheses is at most $\kappa_{\mathrm{tow}}+B_t\leq K_t$. In particular, the dependence on $\tau$ is confined to the required lower bound on $n$.
\end{proof}

\subsection{Partitions in the Final Graph}\label{subsec:final-graph-partitions}

It remains to transfer these partitions to the final graph $X$. For every $2\leq i\leq t$, we project the given set to $X_{i-1}$, apply \Cref{lemma:adjacent-level-partition} with parameter $\tau/2$, and place vertices in the same part whenever their projections lie in the same part. The distance comparison in \Cref{lemma:tower-distance} bounds the resulting diameter. Since projection does not increase distances, the separation bound is preserved.

\begin{proof}[Proof of \Cref{lemma:tower-partition}]
    Let $K_t$ be the constant from \Cref{lemma:adjacent-level-partition}. We fix an integer $i$ with $2\leq i\leq t$ and a set $Z\subseteq V(X)$ with $|Z|\leq n$. The claim is immediate if $Z=\varnothing$, so we assume that $Z$ is nonempty and project it to $Z':=\{\pi_{i-1}(x):x\in Z\}\subseteq V(X_{i-1})$. Since $|Z'|\leq n$, \Cref{lemma:adjacent-level-partition}, applied with parameter $\tau/2$, gives a $(\tau/2)\sigma_i$-bounded random partition $\mathcal P'$ of $Z'$ satisfying
    \[
        \Pr[\mathcal P'(u)\neq\mathcal P'(v)]
        \leq K_t\frac{d_{X_{i-1}}(u,v)}{\sigma_{i-1}}
        \qquad\text{for every }u,v\in Z'.
    \]
    We construct $\mathcal P$ by replacing each part $C\in\mathcal P'$ with $\{x\in Z:\pi_{i-1}(x)\in C\}$.

    Let $x,y$ lie in the same part of $\mathcal P$. Their projections lie in the same part of $\mathcal P'$, so \Cref{lemma:tower-distance} gives
    \[
    \begin{aligned}
        d_X(x,y)
        &\leq
        d_{X_{i-1}}(\pi_{i-1}(x),\pi_{i-1}(y))
        +2\eta\sum_{j=1}^{i-2}\sigma_j\\
        &\leq
        \frac{\tau}{2}\sigma_i+2\eta t\sigma_{i-1}\\
        &\leq\tau\sigma_i.
    \end{aligned}
    \]
    The last inequality holds for sufficiently large $n$, since
    $\sigma_i/\sigma_{i-1}=(\log n)^{1/(t+1)}$
    and $\tau$ is fixed. Thus, $\mathcal P$ is $\tau\sigma_i$-bounded. Finally, for every $x,y\in Z$, the definition of $\mathcal P$ and \Cref{lemma:tower-distance} imply
    \[
    \begin{aligned}
        \Pr[\mathcal P(x)\neq\mathcal P(y)]
        &=
        \Pr[
            \mathcal P'(\pi_{i-1}(x))
            \neq
            \mathcal P'(\pi_{i-1}(y))
        ]\\
        &\leq
        K_t
        \frac{
            d_{X_{i-1}}(\pi_{i-1}(x),\pi_{i-1}(y))
        }{\sigma_{i-1}}\\
        &\leq K_t\frac{d_X(x,y)}{\sigma_{i-1}}.
    \end{aligned}
    \]
\end{proof}

\section{The Integral Lower Bound}\label{sec:integral-lower-bound}

In this section, we prove that every integral assignment for the instance constructed in Section~\ref{sec:gap-instance} has large cost. We fix an integer $t\geq 2$ for the lower-bound argument. Our proof proceeds by contradiction: an assignment of small cost yields a representative close to the labels of most vertices, whereas most vertices themselves are far from it. Their terminal edges then force a large total cost. We establish the following bound.

\begin{lemma}\label{lemma:integral-lower-bound}
    There is a constant $c_t>0$ such that, for every sufficiently large even integer $n$, every integral assignment for the instance $\mathcal{I}$ constructed in Section~\ref{subsec:gap-construction} has cost at least $c_tN\rho=c_tN(\log n)^{(t-1)/(t+1)}$.
\end{lemma}

We choose representatives for $1$-clouds in Section~\ref{subsec:cloud-representatives} and construct representatives for larger clouds in Section~\ref{subsec:parent-representatives}. In Section~\ref{subsec:terminal-cost}, we use the representative of $V(X)$ to bound the terminal-edge costs and complete the proof of \Cref{thm:earthmover-gap}.

\subsection{Representatives and the Base Case}\label{subsec:cloud-representatives}

Let $n$ be a sufficiently large even integer, and let $X$ be the final graph of the valid expander tower used to construct $\mathcal{I}$. We retain $N=n^t$, $\rho=(\log n)^{(t-1)/(t+1)}$, and $A=\sigma_1$ from Section~\ref{subsec:gap-construction}. We identify an integral assignment with a map $f:V(X)\to V(X)$: each vertex $x\in V(X)$ is assigned to the terminal $f(x)_T$, and every terminal is assigned to itself. We call $f(x)$ the \emph{label} of $x$. For every $i\in[t]$, let $E_i$ be the set of level-$i$ edges of $X$. The cost of the assignment is
\begin{equation}\label{eq:integral-cost}
    \begin{aligned}
        \Phi(f)
        &=\frac{2\rho}{\ell_1}
          \sum_{\{x,y\}\in E_1}d_X(f(x),f(y))+\sum_{i=2}^{t}\frac{1}{\ell_i}
          \sum_{\{x,y\}\in E_i}d_X(f(x),f(y))
          +\frac{1}{A}\sum_{x\in V(X)}d_X(x,f(x)).
    \end{aligned}
\end{equation}
The first two terms come from edges of $X$, and the last term comes from terminal edges. We assume for a contradiction that $\Phi(f)<cN\rho$, where $c>0$ will be chosen in Section~\ref{subsec:terminal-cost}.

We first record an elementary consequence of expansion: if a partition of $V(H)$ has few crossing edges, then one part contains most vertices.

\begin{lemma}\label{lemma:expansion-largest-part}
    Let $\mathcal P$ be a partition of $V(H)$, let $M:=\max_{C\in\mathcal P}|C|$, and let $q$ be the number of edges of $H$ whose endpoints belong to different parts. Then, $n-M\leq 2q/h_0$.
\end{lemma}

\begin{proof}
    Let $C^*$ be a largest part. The claim is immediate if $M=n$. If $n/2<M<n$, expansion applied to the complement of $C^*$ gives $q\geq|\delta_H(C^*)|\geq h_0(n-M)$. Otherwise, every part has size at most $n/2$, so we may sum the expansion bound over all parts. Each crossing edge is counted twice, giving $2q=\sum_{C\in\mathcal P}|\delta_H(C)|\geq h_0n$. Both cases imply the claimed bound.
\end{proof}

We construct representatives in increasing order of cloud size. For every cloud $C$, we choose its representative $z_C$ from the labels $\{f(x):x\in C\}$; thus, $z_C$ is a vertex of $X$ but need not belong to $C$. Let $0<\tau\leq\tau_0$, where $\tau_0$ is as in Section~\ref{sec:random-partitions}; we will choose $\tau$ in Section~\ref{subsec:terminal-cost}. We set $r_1:=0$ and $r_i:=r_{i-1}+\tau\sigma_i$ for every $2\leq i\leq t$. Since consecutive scales differ by a factor $(\log n)^{1/(t+1)}\geq 2$ for sufficiently large $n$, these radii satisfy $r_i\leq 2\tau\sigma_i$ for every $i\in[t]$. A vertex $x$ of an $i$-cloud $C$ is \emph{good for $C$} if $d_X(f(x),z_C)\leq r_i$ and \emph{bad for $C$} otherwise. Let $e_C$ be the fraction of vertices of $C$ that are bad for $C$. For every $i\in[t]$, let $\mathcal C_i$ be the collection of $i$-clouds and let $e_i:=|\mathcal C_i|^{-1}\sum_{C\in\mathcal C_i}e_C$ be their average error. The $i$-clouds partition $V(X)$ into sets of size $n^i$, so $Ne_i$ is the total number of bad vertices at level $i$. We construct representatives for every cloud and use the cost assumption to bound these errors.

For the base case, we choose $z_C$ to be a most frequent label in each $1$-cloud $C$. Since $r_1=0$, a vertex is bad precisely when its label differs from $z_C$. Let $q_1$ be the total number of level-$1$ edges whose endpoints have different labels. Each $1$-cloud is a copy of $H$, so applying \Cref{lemma:expansion-largest-part} to its partition by labels and summing over all $1$-clouds gives $Ne_1\leq 2q_1/h_0$. Distinct labels have distance at least $\ell_1$, since every edge of $X$ has length at least $\ell_1$. Each of these $q_1$ edges therefore contributes at least $2\rho$ to $\Phi(f)$. Combining the two bounds yields 
\[
    e_1\leq \frac{\Phi(f)}{h_0N\rho}<\frac{c}{h_0}.
\]

\subsection{Constructing Parent Representatives}\label{subsec:parent-representatives}

We fix an integer $2\leq i\leq t$ and assume that representatives have been chosen for every $(i-1)$-cloud. Let $C$ be an $i$-cloud. We identify its corresponding core with $H$ and, for every $u\in V(H)$, let $C_u$ be the child indexed by $u$. Each child has $n^{i-1}$ vertices, representative $z_u:=z_{C_u}$, and error $e_u:=e_{C_u}$. The set $Z_C:=\{z_u:u\in V(H)\}$ has at most $n$ distinct vertices. By \Cref{lemma:tower-partition}, it admits a $\tau\sigma_i$-bounded random partition $\mathcal P$ with
\[
    \Pr[\mathcal P(z_u)\neq\mathcal P(z_v)]\leq K_t \frac{d_X(z_u,z_v)}{\sigma_{i-1}}
\]
for every $u,v\in V(H)$. We use matching costs to bound the expected number of edges of $H$ whose child representatives lie in different parts. We then group the children according to these parts and use expansion to bound the number outside the largest group.

For every $\{u,v\}\in E(H)$, let $M_{uv}$ be the matching of surviving level-$i$ lifts between $C_u$ and $C_v$. By \Cref{lemma:cloud-matching}, it contains at least $(1-L_n)n^{i-1}$ edges, where $L_n:=(t-1)\lceil n^{0.99}\rceil/n$. Let
\[
    E_{uv}:=\sum_{\{x,y\}\in M_{uv}}d_X(f(x),f(y)),
\]
and $E_{uv}/\ell_i$ is this matching's contribution to $\Phi(f)$. If both children have small errors, many matching edges join vertices whose labels are close to the two representatives. A small matching cost then forces the representatives to be close, as quantified by the following lemma.

\begin{lemma}\label{lemma:representative-separation}
    For every $\{u,v\}\in E(H)$,
    \[
        \Pr[\mathcal P(z_u)\neq\mathcal P(z_v)]
        \leq 2(L_n+e_u+e_v)
        +\frac{2K_t}{\sigma_{i-1}}
         \left(\frac{E_{uv}}{n^{i-1}}+r_{i-1}\right).
    \]
\end{lemma}

\begin{proof}
    If $L_n+e_u+e_v\geq 1/2$, the first term on the right-hand side is at least one, and the claim follows. Otherwise, since $M_{uv}$ is a matching, at most $(e_u+e_v)n^{i-1}$ of its edges have an endpoint that is bad for its child. Hence, more than $n^{i-1}/2$ edges have both endpoints good. For every such edge $\{x,y\}$, with $x\in C_u$ and $y\in C_v$, the triangle inequality gives $d_X(z_u,z_v)\leq d_X(f(x),f(y))+2r_{i-1}$. Averaging over these edges yields 
    \[
        d_X(z_u,z_v)\leq \frac{2E_{uv}}{n^{i-1}}+2r_{i-1}.
    \]
    The separation bound from \Cref{lemma:tower-partition} now proves the claim.
\end{proof}

We choose the parent representative using a partition with few crossing edges. Let $q_C(\mathcal P)$ count the edges $\{u,v\}\in E(H)$ with $\mathcal P(z_u)\neq\mathcal P(z_v)$, and choose a $\tau\sigma_i$-bounded realization $\mathcal P_C$ satisfying $q_C(\mathcal P_C)\leq\mathbb E[q_C(\mathcal P)]$. For every part $B\in\mathcal P_C$, let $I_B:=\{u\in V(H):z_u\in B\}$. These sets partition $V(H)$, and their sizes count children even when several children share a representative. Let $I^*$ be a largest such set. We choose $u^*\in I^*$ and set $z_C:=z_{u^*}$, which is a label attained in $C_{u^*}\subseteq C$. For every $u\in I^*$, the diameter bound gives $d_X(z_u,z_C)\leq\tau\sigma_i$. If $x\in C_u$ is good for its child, then $d_X(f(x),z_C)\leq r_{i-1}+\tau\sigma_i=r_i$, so $x$ is also good for $C$. A bad vertex of $C$ must therefore be bad for its child or belong to a child outside $I^*$. We use this observation to bound the parent error.

\begin{lemma}\label{lemma:parent-error}
    The representative $z_C$ chosen above satisfies
    \[
        e_C\leq
        \left(1+\frac{40}{h_0}\right)\frac{1}{n}\sum_{u\in V(H)}e_u
        +\frac{20}{h_0}L_n
        +\frac{4K_t}{h_0n^i\sigma_{i-1}}\sum_{\{u,v\}\in E(H)}E_{uv}
        +\frac{20K_t}{h_0}\frac{r_{i-1}}{\sigma_{i-1}}.
    \]
\end{lemma}

\begin{proof}
    By \Cref{lemma:expansion-largest-part} and the choice of $\mathcal P_C$, at most $2\mathbb E[q_C(\mathcal P)]/h_0$ children lie outside $I^*$. Since all children have the same size, the preceding observation gives 
    \[
        e_C\leq n^{-1}\sum_{u\in V(H)}e_u+\frac{2\mathbb E[q_C(\mathcal P)]}{h_0n}.
    \]
    To bound the expectation, we sum \Cref{lemma:representative-separation} over the edges of $H$. The maximum degree is at most $10$, so $|E(H)|\leq 5n$, and the sum of $e_u+e_v$ over all edges counts each $e_u$ at most ten times. Consequently,
    \[
        \mathbb E[q_C(\mathcal P)]
        \leq 10nL_n+20\sum_{u\in V(H)}e_u
        +\frac{2K_t}{n^{i-1}\sigma_{i-1}}\sum_{\{u,v\}\in E(H)}E_{uv}
        +\frac{10K_tnr_{i-1}}{\sigma_{i-1}}.
    \]
    Substituting this estimate into the bound for $e_C$ proves the lemma.
\end{proof}

Applying this construction to every $i$-cloud completes the induction at level $i$. We next average the error bound over all clouds at that level, so the matching costs become the total cost of level-$i$ edges. For every $2\leq i\leq t$, let 
\[
    \Phi_i:=\ell_i^{-1}\sum_{\{x,y\}\in E_i}d_X(f(x),f(y))
\]
denote this contribution to $\Phi(f)$.

\begin{lemma}\label{lemma:level-error}
    For every $2\leq i\leq t$,
    \[
        e_i\leq
        \left(1+\frac{40}{h_0}\right)e_{i-1}
        +\frac{20}{h_0}L_n
        +\frac{4K_t}{h_0}\frac{\Phi_i\ell_i}{N\sigma_{i-1}}
        +\frac{20K_t}{h_0}\frac{r_{i-1}}{\sigma_{i-1}}.
    \]
\end{lemma}

\begin{proof}
    We average \Cref{lemma:parent-error} over the $N/n^i$ clouds in $\mathcal C_i$. Every $(i-1)$-cloud is a child of exactly one $i$-cloud, and every parent has $n$ children, so the average of the child-error term is $(1+40/h_0)e_{i-1}$. For every $C\in\mathcal C_i$, let $E_{uv}^C$ denote the matching quantity associated with $C$. Each level-$i$ edge occurs in exactly one of these matchings, giving $\sum_{C\in\mathcal C_i}\sum_{\{u,v\}\in E(H)}E_{uv}^C=\ell_i\Phi_i$. Multiplying by the averaging factor $n^i/N$ therefore gives the term $4K_t\Phi_i\ell_i/(h_0N\sigma_{i-1})$. The terms involving $L_n$ and $r_{i-1}$ are the same for every cloud, which proves the recurrence.
\end{proof}

\subsection{Terminal Edges Force Large Cost}\label{subsec:terminal-cost}

We now iterate the error bound through all $t$ levels. Under the small-cost assumption, the representative of the unique $t$-cloud $V(X)$ is close to the labels of most vertices.

\begin{lemma}\label{lemma:total-error}
    There is a constant $C_t\geq 1$, independent of $c$, $\tau$, and $n$, such that, if $\Phi(f)<cN\rho$, then the representatives constructed above satisfy $e_t\leq C_t(c+\tau+L_n)$.
\end{lemma}

\begin{proof}
    For every $2\leq i\leq t$, we have $\Phi_i\leq\Phi(f)<cN\rho$. The identity $\rho\ell_i/\sigma_{i-1}=(\log n)^{-1/(t+1)}$ therefore gives $\Phi_i\ell_i/(N\sigma_{i-1})\leq c$ for sufficiently large $n$. We also have $r_{i-1}/\sigma_{i-1}\leq 2\tau$. Let $\lambda:=1+40/h_0$. Since $K_t\geq 1$, \Cref{lemma:level-error} yields
    \[
        e_i\leq\lambda e_{i-1}+\frac{40K_t}{h_0}(c+\tau+L_n).
    \]
    Starting from the base-case bound $e_1<c/h_0$ and iterating this recurrence $t-1$ times proves the lemma for a suitable constant $C_t\geq 1$. This constant depends only on $t$, $h_0$, and $K_t$, all of which are independent of $c$, $\tau$, and $n$.
\end{proof}

We complete the integral lower bound by comparing the labels with the original vertices. The preceding lemma places most labels near one representative. On the other hand, projection onto the bounded-degree graph $X_t=H$ shows that only a small fraction of the vertices of $X$ can lie near that representative. Many vertices must therefore be far from their labels, forcing a large cost on their terminal edges.

\begin{proof}[Proof of \Cref{lemma:integral-lower-bound}]
    We choose positive constants $\tau$ and $c$ such that 
    \[
        \tau\leq\min\{\tau_0,1/(16C_t),1/(16\log 20)\} \qquad \text{and} \qquad c\leq\min\{1/(16C_t),1/(32\log 20)\}.
    \]
    We take $n$ sufficiently large for all preceding bounds and to ensure $C_tL_n\leq 1/16$ and $20n^{-3/4}\leq 1/4$. These choices depend only on $t$. Under the assumption $\Phi(f)<cN\rho$, \Cref{lemma:total-error} gives $e_t\leq 3/16<1/4$. Let $z$ be the representative of $V(X)$, and set $R:=\sigma_t/(4\log 20)$. At least $3N/4$ vertices $x\in V(X)$ satisfy $d_X(f(x),z)\leq r_t\leq 2\tau\sigma_t\leq R/2$.

    To bound the number of vertices close to $z$, we consider the ball $B:=B_{X_t}(\pi_t(z),R)$. Every edge of $X_t=H$ has length $\ell_t$, so every vertex of $B$ is reachable from $\pi_t(z)$ by a path with at most $R/\ell_t=\log n/(4\log 20)$ edges. Since $H$ has maximum degree at most $10$, we have 
    \[
        |B|\leq\sum_{j=0}^{\lfloor R/\ell_t\rfloor}10^j\leq 20n^{1/4}.
    \]
    By \Cref{lemma:tower-distance}, projection does not increase distances, so every vertex of $B_X(z,R)$ projects into $B$. Each vertex of $X_t$ has exactly $n^{t-1}$ preimages, giving
    \[
        |B_X(z,R)|\leq n^{t-1}|B|\leq 20Nn^{-3/4}\leq N/4.
    \]

    Thus, at least $N/2$ vertices $x$ lie outside $B_X(z,R)$ while satisfying $d_X(f(x),z)\leq R/2$. For every such vertex, the triangle inequality gives $d_X(x,f(x))\geq d_X(x,z)-d_X(f(x),z)\geq R/2$. Each terminal edge has weight $1/A$, so their contribution yields
    \[
        \Phi(f)\geq\frac{NR}{4A}
        =\frac{N\sigma_t}{16A\log 20}
        =\frac{N\rho}{16\log 20},
    \]
    where the last equality is because $\sigma_t/A=\rho$. This contradicts $\Phi(f)<cN\rho$, since $c\leq 1/(32\log 20)$. Taking $c_t:=c$ proves the lemma.
\end{proof}

Combining this integral lower bound with the fractional upper bound from Section~\ref{sec:gap-instance} completes the proof of our main integrality-gap result.

\begin{proof}[Proof of \Cref{thm:earthmover-gap}]
    We fix $\varepsilon\in(0,1)$ and choose an integer $t\geq 2$ such that $2/(t+1)\leq\varepsilon$. Let $p:=(t-1)/(t+1)$, so $p\geq 1-\varepsilon$. For every sufficiently large even integer $n$, the instance $\mathcal{I}$ from Section~\ref{subsec:gap-construction} has $k=N=n^t$ terminals. By \Cref{lemma:fractional-solution,lemma:integral-lower-bound},
    \[
        \frac{\operatorname{OPT}(\mathcal{I})}{\operatorname{EM}(\mathcal{I})}
        \geq\frac{c_t}{\eta+10t}(\log n)^p
        =\frac{c_t}{(\eta+10t)t^p}(\log k)^p
        \geq c_\varepsilon(\log k)^{1-\varepsilon},
    \]
    where $c_\varepsilon:=c_t/((\eta+10t)t^p)>0$ depends only on $\varepsilon$. We recall that $k=n^t$ can be arbitrarily large, which finishes the proof.
\end{proof}

\paragraph{AI Methodology.} 
GPT-6 Astra initially derived an almost tight integrality gap for the metric relaxation of the $0$-Extension problem. The authors then proposed extending this result to the earthmover relaxation and developed the proof presented in this work with assistance from GPT-6 Astra. The model also assisted with drafting and revising the manuscript. The authors verified the correctness and originality of the results, including references, and take full responsibility for the content of this work.

\bibliographystyle{alpha}
\bibliography{references}

\end{document}